\documentclass{aa}  

\usepackage{natbib}
\usepackage{graphicx}
\usepackage{amsmath}
\usepackage{mhchem}
\usepackage{xcolor}
\usepackage{ulem} 
\usepackage{caption}
\usepackage{float}

\usepackage[colorlinks=true, allcolors=blue]{hyperref}
\usepackage{soul}

\usepackage{subcaption}         
\usepackage{lscape}             
\usepackage{placeins}           

\usepackage{txfonts}
\begin{document}

   \title{Chemical constraints on the dynamical evolution of the cold core L694 \thanks{This work is based on observations carried out under project number ID 111-21 with the 30m telescope. IRAM is supported by INSU/CNRS (France), MPG (Germany) and IGN (Spain).  }
 }

   \author{A. Taillard,
          \inst{1}
          V. Wakelam \inst{2}, P. Gratier \inst{2}, E. Dartois \inst{3}, M. Chabot \inst{4}, J. A. Noble \inst{5}, L. Chu \inst{6}
          }

  \institute{Centro de Astrobiología (CAB), CSIC-INTA, Ctra. de Ajalvir, km 4, Torrejón de Ardoz, 28850 Madrid, Spain
              \email{angele.taillard@cab.inta-csic.es}
         \and
             Laboratoire d'Astrophysique de Bordeaux (LAB), Univ. Bordeaux, CNRS, B18N, allée Geoffroy Saint-Hilaire, 33615 Pessac, France
         \and
             Institut des sciences Moléculaires d'Orsay, CNRS, Université Paris-Saclay, Bât 520, Rue André Rivière, 91405 Orsay, France
         \and
             Université Paris-Saclay, CNRS/IN2P3, IJCLab, 91405 Orsay, France
         \and 
             Physique des Interactions Ioniques et Moléculaires, CNRS, Aix Marseille Univ., 13397 Marseille, France
         \and
            Institute for Astronomy, 2680 Woodlawn Drive, Honolulu, HI 96822-1897, USA
             }

   \date{To be submitted}

 
  \abstract
   {
    In star-forming regions, molecular cloud history and dynamics set the trend in the chemical composition. Ice formation, in particular, is affected by the evolution of physical conditions, which can lead to different ice compositions within the same cloud. In cold cores with medium densities $>$ 10$^4$ cm$^{-3}$, low temperatures $<$ 15 K, and low UV radiation $<$ G$_0$, most complex organic molecules are formed on dust grain surfaces and are released back into the gas phase through non-thermal mechanisms such as sputtering or heating by cosmic-rays, photodesorption, or chemical desorption. Studying both gas- and solid-phases can help observers to add constraints on the chemical and dynamical evolution of cold cores. \\
    We present a study of the cold core L694, observed with the IRAM 30m single-dish radio telescope. Observed species include CO (and its isotopologues) and CH$_3$OH, a key chemical species precursor of more complex organic molecules. \\
    We applied an inverted non-local thermal equilibrium radiative transfer code, previously used on observations of the pre-stellar core L429-C, in order to obtain gas-phase abundances by deriving the column densities of the detected species from the spectroscopic parameters of the targeted molecular transitions (intensity, line width), and from physical parameters derived from archival observations (H$_2$ volume density and gas temperature). This allowed us to probe the molecular abundances as a function of density and visual extinction. 
    In parallel, we ran chemical models (both static and dynamic) to constrain the evolution time of the core by directly comparing the observations with the model outputs.
    We then compared the compositions of the cold cores L429-C and L694. \\
    The gas-phase abundances in L694 all exhibit a common depletion profile (with high variability in the depletion factor), as the core is identified to be in a more advanced (infalling) state compared to L429-C. The physical parameters of the two cores are, however, very similar, leading to close evolutionary timescales in our static models. The dynamical model fails to reproduce the CO gas-phase abundances at high density, predicting an evolutionary timescale that is too short compared to static models. A more detailed study on the parameter constraining the CO freeze-out could help to better constrain the timescale.  

}


       \titlerunning{Constraining the dynamical evolution of dense cores with molecular abundances}
   \authorrunning{Taillard et al.}

   \maketitle
%

\section{Introduction}

Complex organic molecules \citep[COMs, composed of six atoms or more, as defined by][]{herbst_complex_2009} begin to form in the early phases of star formation.
These steps are crucial for setting the trend in the chemical composition for the protostellar environment. Cold cores are dependent on the physical conditions of their original molecular cloud. They are characterised by densities higher than 10$^{3}$ cm$^{-3}$, temperatures lower than 20 K, and visual extinctions higher than 3 mag \citep{2007ARA&A..45..339B}. Since the temperature is low, the chemistry in the gas phase is limited, while a rich chemistry takes place at the surface of interstellar dust grains. Grains act as catalysts and play the role of the 'meeting point' for atoms, radicals, and other molecules: 
species at the surface diffuse and react with other molecules to form more complex products \citep{cuppen_grain_2017}. In translucent and dense clouds with Av $>$ 1.5, the grains are covered by a mantle where H$_2$O ice is the dominant component \citep{boogert_observations_2015}.
A shift occurs at densities $> 10^5$ cm$^{-3}$, where volatile CO efficiently freezes out onto the grains \citep{bacmann_degree_2002,qasim_formation_2018}. The CO freeze-out leads to the formation of COMs on short timescales of a few tens of thousands of years \citep{2009A&A...508..275C}, with the build-up of a CO-dominated layer over the H$_2$O-dominated mantle. 
Infrared observational studies have revealed cold core regions presenting high column densities of COMs, such as CH$_3$OH \citep{boogert_ice_2011,chu_observations_2020,perotti_linking_2020,mcclure_ice_2023}.

These regions are of particular interest, since they show a large variability in molecular composition despite sharing similar physical conditions. For instance, this variety can be found within a single source \citep[e.g. in ices,][]{noble_two-dimensional_2017}. This is the case in the Taurus Molecular Cloud (TMC), located at a distance that varies between $\sim$ 128 and 198 pc depending on the part of the cloud being studied \citep{2002ApJ...575..950O,2019A&A...630A.137G}. Within the cloud, the widely targeted filament TMC-1 \citep[and reference therein]{2013ChRv..113.8710A} exhibits different chemical behaviour at different locations. Over a wider group of targets, \citet{2019A&A...624A.105F} showed that for the same density of $\sim$ 10$^5$ cm$^{-3}$ in various clouds, the CS gas-phase abundance varies by over more than one order of magnitude. 
Studying such differences through chemical modelling, which employs similar physical parameter inputs, can help convey the most probable hypothesis for each cloud. Yet, the chemistry within a model can be completely affected by a single parameter that may not be well constrained.

In this work, we studied the chemical composition of the cold core L694 using gas-phase observations obtained with the IRAM 30m telescope and through chemical modelling. 
By comparing observed and modelled abundances, we aimed to put constraints on the formation timescale of cold cores. This paper follows a methodology used in a previous study of the chemical composition of the cold core L429-C \citep{taillard_constraints_2023}. The source, the gas-phase observations, and the observational results are described in Sect.~\ref{Observations}. 
In Sect.~\ref{section-abundances}, we present and discuss the molecular abundances derived from observations. 
The chemical models--both static and dynamic--as well as the comparison of their results with the observations are given in Sect.~\ref{section-models}. Lastly, in Sect.~\ref{section-discussion}, we compare our results from these observations in L694 with those from our previous study on the cold core L429-C  and discuss the time estimation derived from our chemical models.

\section{Source description and new observations}\label{Observations}

\subsection{The LDN 694 cold core}

\begin{figure*}[!htbp] 
    \centering
    \includegraphics[width=0.95\linewidth]{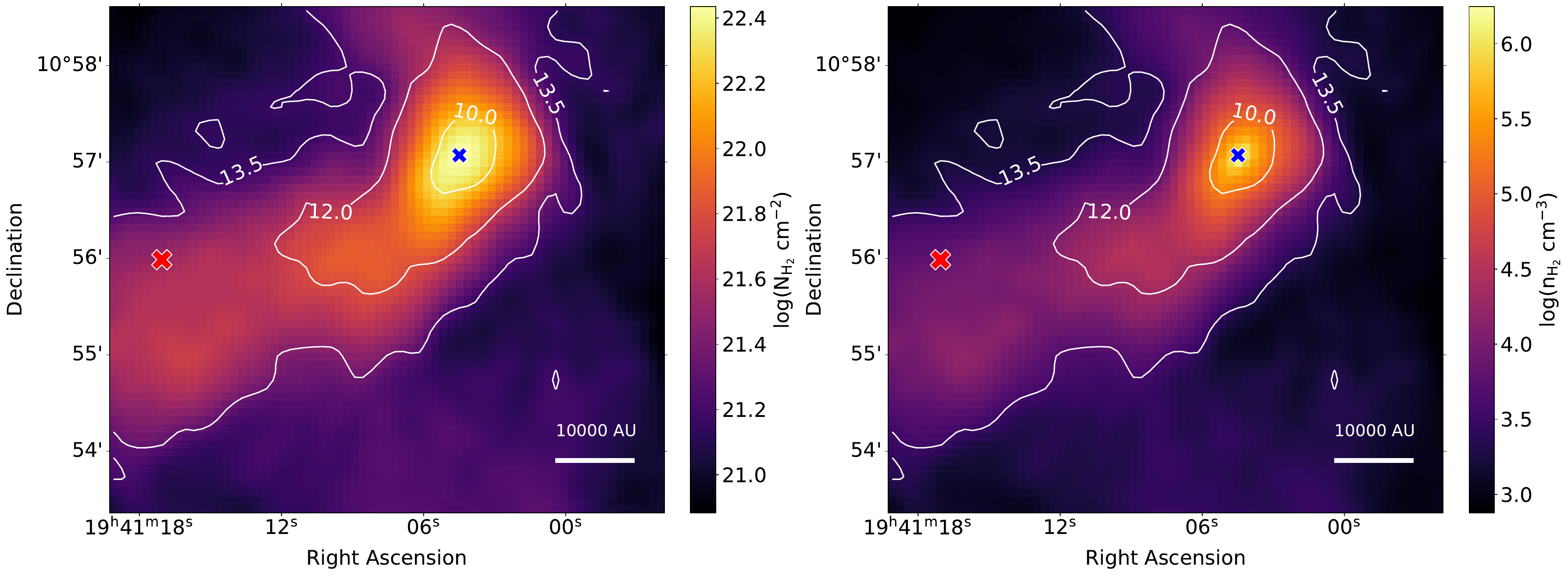}
     \caption{Left: H$_2$ column density map (N$_{\rm H_2}$ in cm$^{-2}$) of L694, computed from \textit{Herschel} maps at 250, 350, and 500 $\mu$m. The position of the continuum peak is represented by the blue cross. The red cross is at an off-position, used to show the different velocity profile in Fig.~\ref{fig:l694-c18o-spectra}. Right: H$_2$ volume density (n$_{\rm H_2}$ in cm$^{-3}$) obtained from the method presented in \citet{marchal_rohsa_2019}. In both maps, the temperature (in K) appears in white contours, and the spatial scale is represented by the horizontal white bar (10 000 AU).}
    \label{fig:L694_herschel_nh2}
\end{figure*}

LDN 694 (sometimes referred to as L694-2, but hereafter L694) is a cold (T $<$ 18 K), dense (n$\rm_{H_2} > 1 \times 10^6$ cm$^{-3}$), and isolated core in the Aquila Rift, located at a distance of approximately 205 pc \citep{kim_role_2022}. 
The core has been labelled as infalling, as molecular line profiles exhibit redshifted self-absorption--strong evidence of inwards motions \citep{lee_survey_2001}--as well as an indication of a velocity gradient, possibly due to rotation \citep{lee_survey_2004} and faster infall motions with supersonic speeds \citep{2007ApJ...660.1326L,2013ApJ...769...50S}. Another study of 
the N$_2$H$^+$ emission shows a centrally peaked dense core, with a line width increasing towards the centre, possibly due to infall \citep{crapsi_probing_2005}.
The N$_2$H$^+$, CS, and DCO$^+$ spectra shown in this above studies present blue-shifted components on the source position.
The core is centrally condensed, with an estimated mass of 1 M$_{\odot}$ \citep{lee_survey_2001}, and is expected to form a protostellar object within 10$^4$ years \citep{2002AJ....124.2756V,williams_highresolution_2006}. No radio point source has been detected so far \citep{2002AJ....123.3325H}, and the core presents a steep density profile that drops quickly (over 0.1 pc) as L694 exhibits an elongated, filament-like structure \citep{chu_observations_2020,kim_role_2022} (see Fig.~\ref{fig:L694_herschel_nh2}). 
In \citet{chu_observations_2020}, the authors deduced a visual extinction up to $\sim$ 27 mag (from the JHK photometric bands) towards the reddest of three detected stars, located at less than 0.1 pc from the centre of the core.

The core presents an advanced chemistry: it has a large column density of ortho-N$_2$D$^+$ \citep[3.2 $\times$ 10$^{13}$ cm$^{-2}$]{caselli_survey_2008} and a high deuterium fractionation (N(N$_2$H$^+$)/N(N$_2$D$^+$) $\sim$ 0.26) \citep{crapsi_probing_2005}. Studies of C$^{34}$S by \citet{kim_role_2022} show that the signal is not centrally peaked. Moreover, the emission distribution suggests high depletion in the high-density region. The core also exhibits a CO depletion (with a depletion factor of 11 to 16) that is larger towards the nucleus \citep{wirstrom_search_2016}.
A particularity of L694 is that the ice observations made by \citet{chu_observations_2020} using the NASA infrared telescope facility (IRTF) show strong infrared absorption at a distance of 0.1 pc from the peak density towards at least one background star near the edge of the core.
The authors measured a CH$_3$OH/H$_2$O ratio of 14.7\% \citep[larger than in L429-C for similar observations,][]{boogert_ice_2011}. Moreover, only a small amount of CO is frozen onto the dust grain ($\leqslant$ 15\% with respect to CO$\rm_{gas}$), but $\sim$ 30\% of the CO ice is mixed with methanol--the highest such ratio measured in their core sample. This high ratio indicates that the mixture traces the densest parts of the cloud, where the CO depletion is the highest and the conversion from CO to CH$_3$OH occurs at a faster rate. L694 also presents a remarkably high CH$_3$OH$\rm_{ice}$/CO$\rm_{tot}$ ratio (0.064) with respect to the entire sample, which is a mixture of background stars and YSO observations within Barnard 59, LDN 483, LDN 694, and Barnard 335. 
In a follow-up study of five molecular cores \citep{chu_constraining_2021}, the authors measured the total hydrogen volume densities to constrain ice formation processes within the same sample and derived Av maps using NIR photometry of a large population of background stars (i.e. L694 data used below in Sect.~\ref{annexe-Chu}). This study shows that the density required to form CH$_3$OH is only reached in less than 2\% of the total core in the densest region.

\subsection{New IRAM 30m observations}

During November 2021, the core was observed for 23 hours with IRAM 30m to provide short-spacing data.
We centred the map at R.A. = 19\textsuperscript{h}41\textsuperscript{m}11\textsuperscript{s}.63, DEC. = -10°56'27.7$\arcsec$ (J2000). Our observation setup focused on four frequency bands corresponding to: 93 - 94.8 GHz, 96.2 - 98.1 GHz, 108.6 - 110.4 GHz, and 111.9 - 113.7 GHz.
With a beam of $\sim$ 28.5$\arcsec$, we obtained $\sim$ $300\arcsec \times 300\arcsec$ maps. 
Our previous study on a different cold core, which included interferometer data (NOEMA), showed no detectable small scale emission in L429-C \citep{taillard_constraints_2023}. We did not expect the beam dilution and the sub-beam source structure to have a significant effect on our results for L694.
Each cube contained 70 velocity channels of 0.15 km.s$^{-1}$. On average, we observed noise sensitivity varying between 0.15 to 0.30~K depending on the molecule. 
The data reduction was performed using the Gildas package CLASS\footnote{\url{http://www.iram.fr/IRAMFR/GILDAS}}. 
We used the Cologne Database for Molecular Spectroscopy\footnote{\url{https://cdms.astro.uni-koeln.de/classic/}}  \citep[CDMS,][]{muller_cologne_2001} and the Jet Propulsion Laboratory catalogue\footnote{\url{https://spec.jpl.nasa.gov}} to conduct line identification. Moreover, we detected all the molecules (and their lines) listed in Table~\ref{tab.molecules}. As the setup differed slightly from our previous study, some molecules (i.e., N$_2$H$^+$, NH$_2$D, and HNCO) were newly observed, while others (CCS, HC$_3$N, CN, and H$_2$S) were not. The molecules targeted give us constraints on the chemical evolution timescale (mainly CO and its isotopologues--CS and CH$_3$OH) and the dynamical state of the cloud (N$_2$H$^+$, CO, and its isotopologues).

\begin{table*}
\caption{\label{tab.molecules} Detected lines and associated spectroscopic information.}
\begin{center}
\begin{tabular}[t]{ l c  c  c c c  }
\hline
\hline
       Molecule  &      Frequency (MHz)  & Transition                  & E$_{\rm up}$ (K) & g$_{\rm up}$ & A$_{\rm ij}$ (s$^{-1}$)  \\
\hline
  $\rm N_2H^+$  &       93173.9         &  (1-0)                       &  4.47   & 15  & $3.62\times 10^{-5}$  \\
  $\rm CH_3OH$  &       96739.3         &  (2-1)                       &  12.5   & 5   & $2.55\times 10^{-6}$  \\
  $\rm CH_3OH$  &       96741.3         &  (2,0)-(1,0)                 &  7      & 5   & $3.40\times 10^{-6}$   \\
  $\rm ^{34}$SO  &       97715.3         &  (2,3)-(1,2)                 &  9.1    & 7   & $1.07\times 10^{-5}$   \\
  CS            &       97980.9         &  (2-1)                       &  7.1    & 5   & $1.67\times 10^{-5}$    \\
  SO            &       109252.2        &  (3,2)-(2-1)                 &  21.1   & 5   & $1.08\times 10^{-5}$   \\
  $\rm C^{18}O$ &       109782.1        &  (1-0)                       &  5.27   & 3   & $6.26\times 10^{-8}$   \\
  HNCO          &       109905.7        &  (5,0,5)-(4,0,4)             &  15.82  & 11  & $1.80\times 10^{-8}$   \\
  NH$_2$D       &       109782.1        &  (1-0)                       &  5.27   & 3   & $6.26\times 10^{-8}$   \\
  $^{13}$CO     &       110201.3        &  (1-0)                       &  5.29   & 3   & $6.29\times 10^{-8}$   \\
  C$^{17}$O     &       112358.7        &  (1-0)                       &  5.39   & 3   & $6.69\times 10^{-8}$   \\
\hline
\end{tabular}
\end{center}
\end{table*}

\subsubsection{Detected lines and intensity maps}

In the frequency band targeted, we detect a total of 11 lines (24, if we considered the hyperfine structure of N$_2$H$^+$) coming from ten different molecules. We considered a line detected when its intensity was three times the mean noise level in the data cube. The following molecules we observed did not fulfil this condition: CCS (6-7) at 93870 MHz, S$^{18}$O (1-2) at 93267 MHz, OCS (8-7) at 97301 MHz, and C$^{34}$S (2-1) at 96412 MHz.
The integrated intensity maps were built from the data cubes using the Spectral Cube python package \citep{ginsburg_radio-astro-toolsspectral-cube_2019}. Molecular emission can be divided into two categories: (1) emission widely observed across the filament (C$^{18}$O, $^{34}$SO, CH$_3$OH, and CS), and (2) emission centred on the continuum peak (HNCO, NH$_2$D, N$_2$H$^+$, and SO). Integrated intensity maps can be found in Appendix~\ref{annexe-cartesintensité} (Fig.~\ref{fig:L694_intensity}), with the position of the continuum peak marked by a blue cross. In the following subsection, we use the cardinal point convention with respect to right ascension to describe the positions on the map.

\subsubsection{Kinematic analysis \& infall}

\begin{figure}[!htb] 
     \centering
     \includegraphics[width=0.7\linewidth]{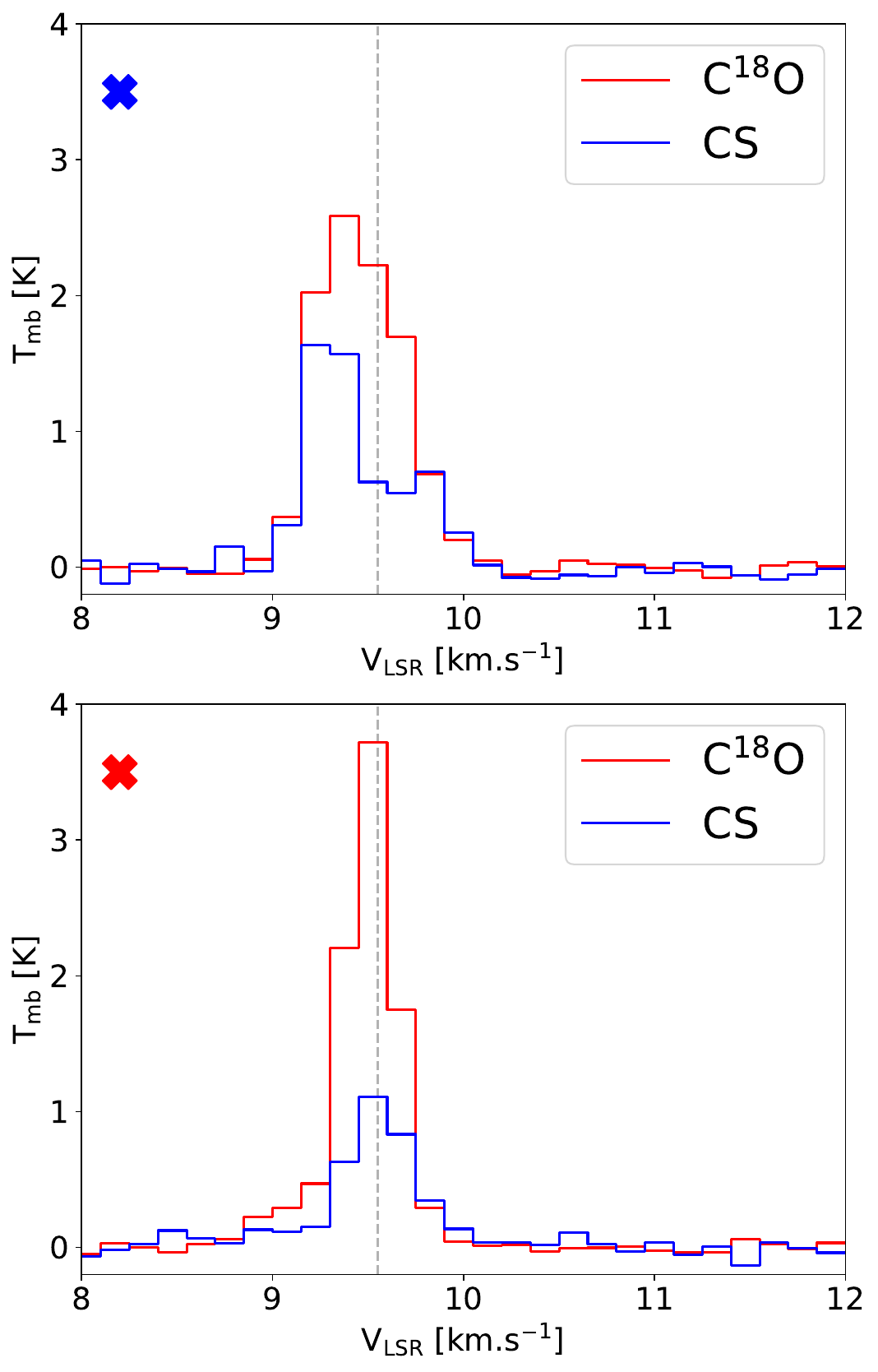}
      \caption{Upper: C$^{18}$O and CS spectra on the core position represented by the dark blue cross on the maps. Lower: Spectra of the same molecules at the position marked by the red cross in Fig~\ref{fig:L694_herschel_nh2}. The grey dashed line is the centroid velocity.
      \label{fig:l694-c18o-spectra} }
 \end{figure}

Most molecules present simple emission profiles that can be fitted by a single Gaussian. The channel velocity maps can be found in Appendix~\ref{velocity_channel_maps} (Figs.~\ref{fig:l694-c18o-channel-maps}). Figure~\ref{fig:l694-c18o-spectra} shows the C$^{18}$O and CS spectra at two positions on the map: the dust continuum peak emission and the off-position in the eastern portion of the map (blue and red crosses, respectively, in Fig.\ref{fig:L694_herschel_nh2}).  
C$^{18}$O presents a Gaussian profile over most of the cloud, with stronger intensity along the filament. Two distinct regions exhibit a deformed Gaussian profile:  on the lower border of the cloud, the red wing is slightly stronger than the blue one, while around the continuum peak, the Gaussian profile flattens, and the blue component becomes stronger (see Fig.~\ref{fig:l694-c18o-spectra}). As discussed in the next section, this coincides with the depletion profile observed in abundance. 
The CS spectrum at the continuum peak exhibits a bimodal profile with a strong component (between 8.9 and 9.6 km.s$^{-1}$) and a weaker component (between 9.6 and 10.4 km.s$^{-1}$). The bimodal profile is evident throughout the entire filament, with the second component weakening until a high intensity peak is reached in the eastern part of the cloud (red cross in Fig.~\ref{fig:L694_herschel_nh2}), where the CS profile becomes a single Gaussian.
This velocity structure is observed in infalling cores such as L1544 \citep{1998ApJ...504..900T}. The authors hypothesized that this signature is due to inward motions towards the core. It was also observed towards a similar region, the TMC, by \citet{lee_survey_2001}. 
L694 was previously labelled as a 'strong infall candidate' in \citet{lee_survey_2004}, where a strong blue component and a less intense red component in the (3-2) CS transition at 146.96 GHz were observed at the position of the source.
In both studies, the authors observed that N$_2$H$^+$ emission is typically found in a significantly more compact region of the cloud compared to the more diffuse CS emission. In the region traced by N$_2$H$^+$, CS exhibits a double-peaked profile, meaning that these two profiles show inward motions.
In our case, the N$_2$H$^+$ intensity is strongest at the continuum peak position (see Fig.~\ref{fig:L694_intensity}). CS, moreover, shows a double-peaked profile. 
All these elements combined confirm the infalling nature of L694.

\subsection{Physical structure of the source and data at hand }\label{sec:physical_param}

We used two methods to investigate the physical parameters of the source based on observations, called method one and method two. 
The data used in method one to derive the physical structure of L694 were taken from the \textit{Herschel} Science Archive\footnote{https://archives.esac.esa.int/} (OBSIDS: 1342230846). L694 was observed with SPIRE at 250, 350, and 500 $\mu$m. 
Different studies estimated the core temperature with values ranging from 8 to 14 K \citep{evans_ii_tracing_2001,harvey_inner_2003,williams_highresolution_2006,seo_internal_2013,chu_constraining_2021,Lin2023}. Using the same pipeline as described in \citet{taillard_constraints_2023}, we derived the dust temperature, the H$_2$ column density, and the H$_2$ volume density across the entire region (Fig.~\ref{fig:L694_herschel_nh2}) observed with the IRAM 30m telescope. We obtained a dust temperature ranging from 9 K at the dust continuum peak to $\sim$ 15 K at the core edges. In Fig.~\ref{fig:l694-parameters} (left), we plot the volume density as a function of the Av obtained from the \textit{Herschel} data. 

Method one follows the standard approach of balancing radiative heating and cooling of grains, as used in numerous studies \citep{2010A&A...518L..87S,2013A&A...551A..98L,2016A&A...592A..61L,sadavoy_intensity-corrected_2018}. 
However, it only probes a certain grain size population, whereas larger grains are known to be cooler and smaller ones warmer \citep{2006PNAS..10312257H}. 
For simplicity, it is often assumed that grains have the same size, typically adopting the canonical grain radius of 0.1 $\mu$m, which represents the mass-weighted average over the grain size distributions \citep{1977ApJ...217..425M,kruegel_physics_2003}). Dust emission can be fitted by spectral energy distribution (SED), and one can deduce a temperature approximation from a modified blackbody model.

We used a second method ('method two') to check if the temperature derived with method one accurately traced the core temperature. We describe the process in Appendix~\ref{appendix_phy_param} and the resulting abundances in Appendix~\ref{annexe-Chu}, using background star photometric measurements and the Av determination method explained in \citet{chu_constraining_2021} for L694. We then converted the Av to N$\rm_{H_2}$. Based on the method described in \citet{hocuk_parameterizing_2017}, we estimated T$_{dust}$ using the H$_2$ column density.
We computed the temperature, Av, and H$_2$ column density, shown in the central panel of Fig.~\ref{fig:l694-parameters}. 
The temperature obtained is, overall, 3 K lower over the density grid, ranging from 7 to 12 K. Compared to method one, Av increases less steadily but covers the same range of values (2 to 30 mag). The density is slightly higher by a factor of 2 towards the edge of the cloud but is the same at the continuum dust peak position. The main issue with method two is that it does not probe the deepest region of the cloud. Therefore, we rely on the kernel density estimation to smooth the values around the dust continuum peak \citep[see][for more details]{chu_constraining_2021}.
Overall, this method gives us almost the same range of values for Av and the H$_2$ column density as in method one, but gives a slightly lower temperature.
The physical conditions derived from these two methods are used to study the chemistry discussed in Sect.~\ref{section-abundances} and Appendix~\ref{annexe-Chu}.

\section{Chemical composition of L694}\label{section-abundances}

\subsection{Molecular abundance maps}

In order to obtain molecular abundance (X$\rm_{mol}$ = N$\rm_{mol}$/N$\rm_{H_2}$) maps from our gas-phase observations, we used the method described in \citet{taillard_constraints_2023}, employing \textit{Herschel} maps of the regions to derive the physical parameters (i.e. method one). The isotopic ratio we used for the conversion between C$^{18}$O and CO is 557 \citep{wilson_isotopes_1999}.
Each molecule was treated using ndRADEX \citep{taniguchi_astropenguinndradex_2020}, with collision rates extracted from the following original studies: NH$_2$D from \citet{muller_cologne_2001}, N$_2$H$^+$ from \citet{schoier_atomic_2005}, SO from \citet{lique_rotational_2005}, CS from \citet{lique_rotational_2006}, CO from \citet{yang_rotational_2010}, CH$_3$OH  from \citet{rabli_rotational_2010}, and HNCO from \citet{sahnoun2018}.

\begin{figure*} 
    \centering
    \includegraphics[width=0.44\linewidth]{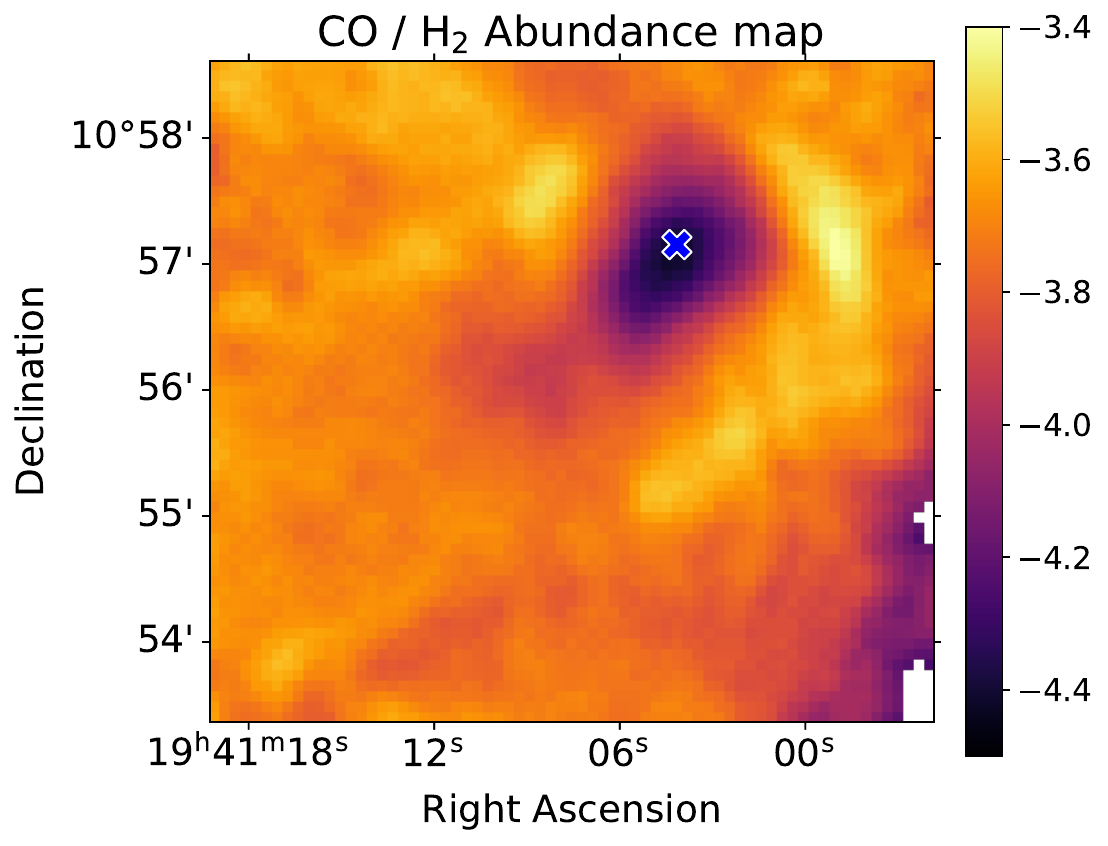}
    \includegraphics[width=0.45\linewidth]{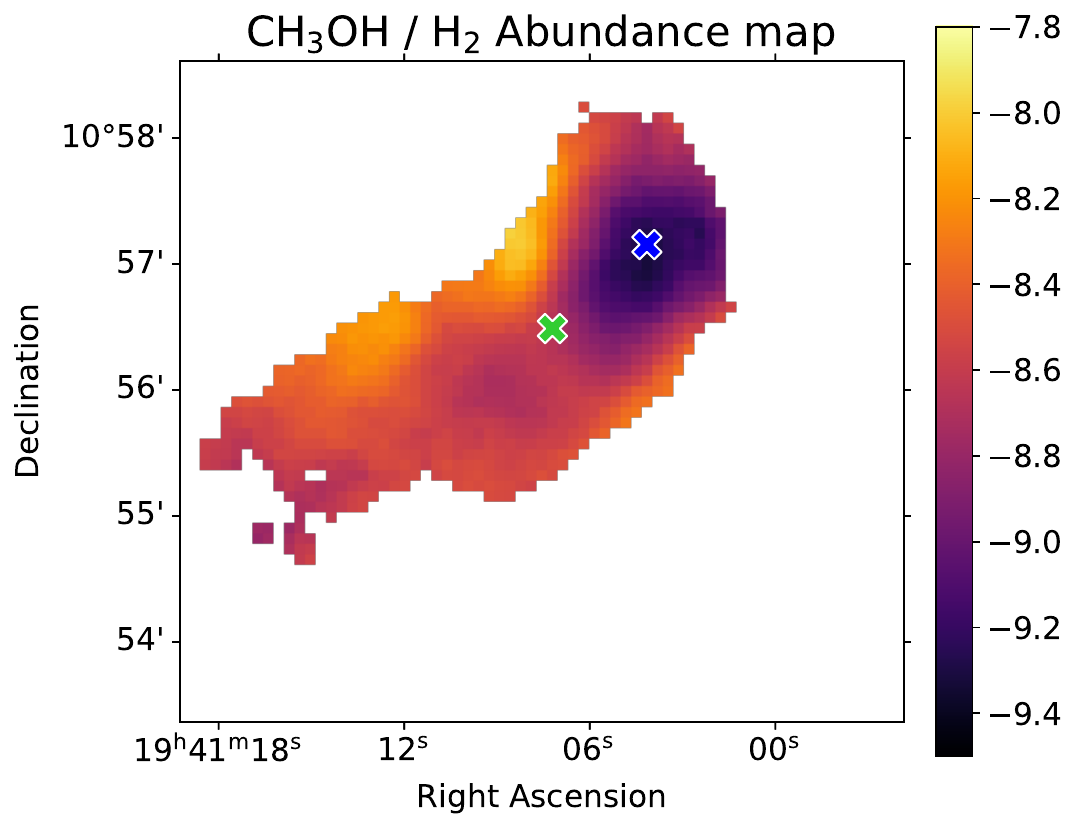}
    \includegraphics[width=0.44\linewidth]{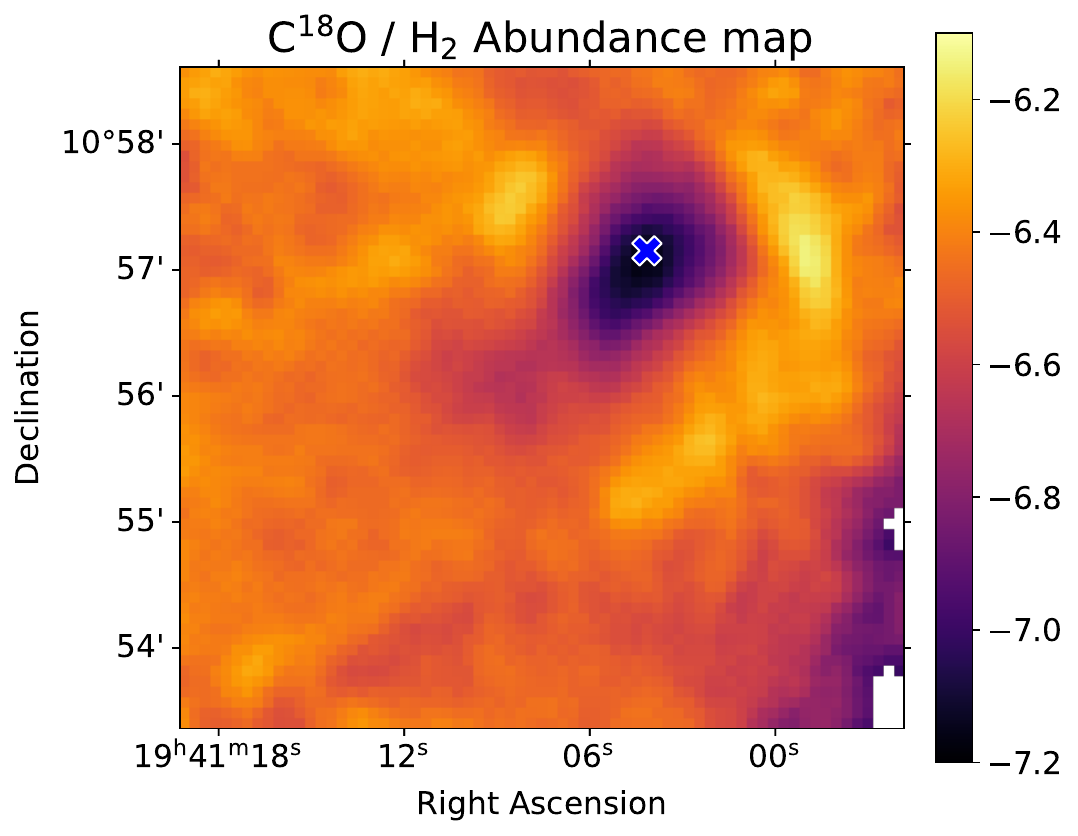}
    \includegraphics[width=0.45\linewidth]{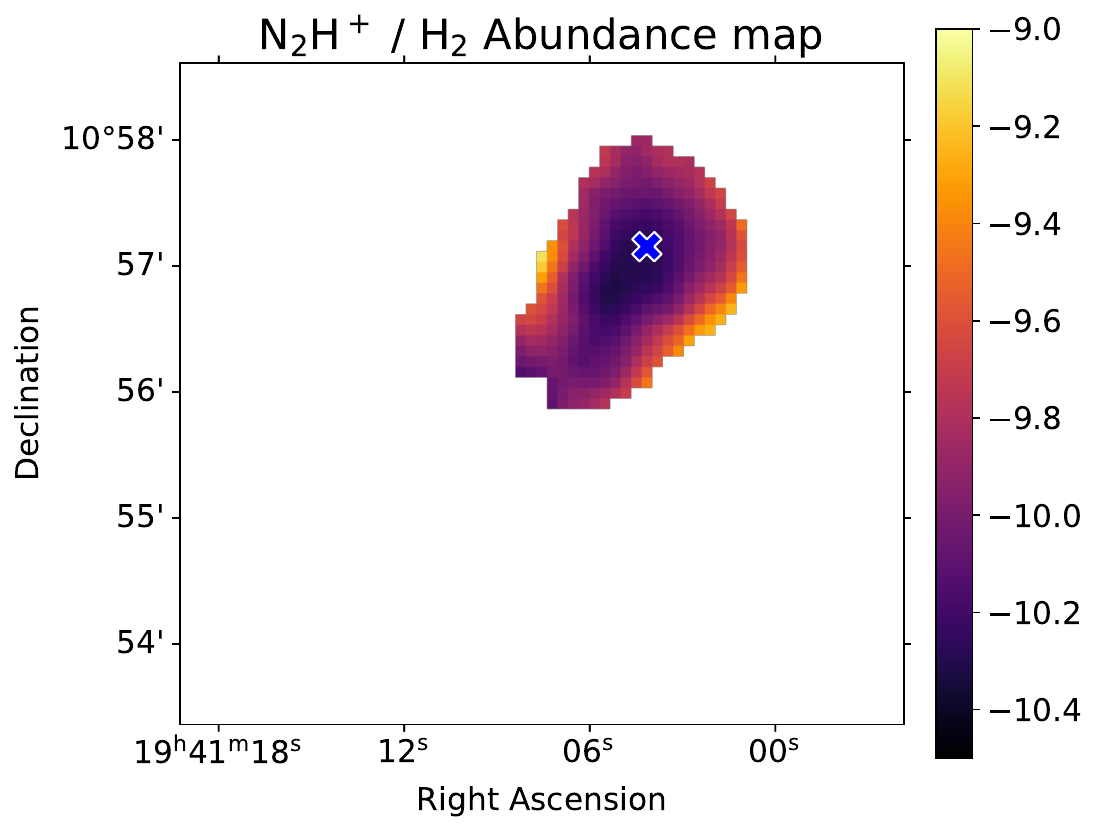}
    \includegraphics[width=0.45\linewidth]{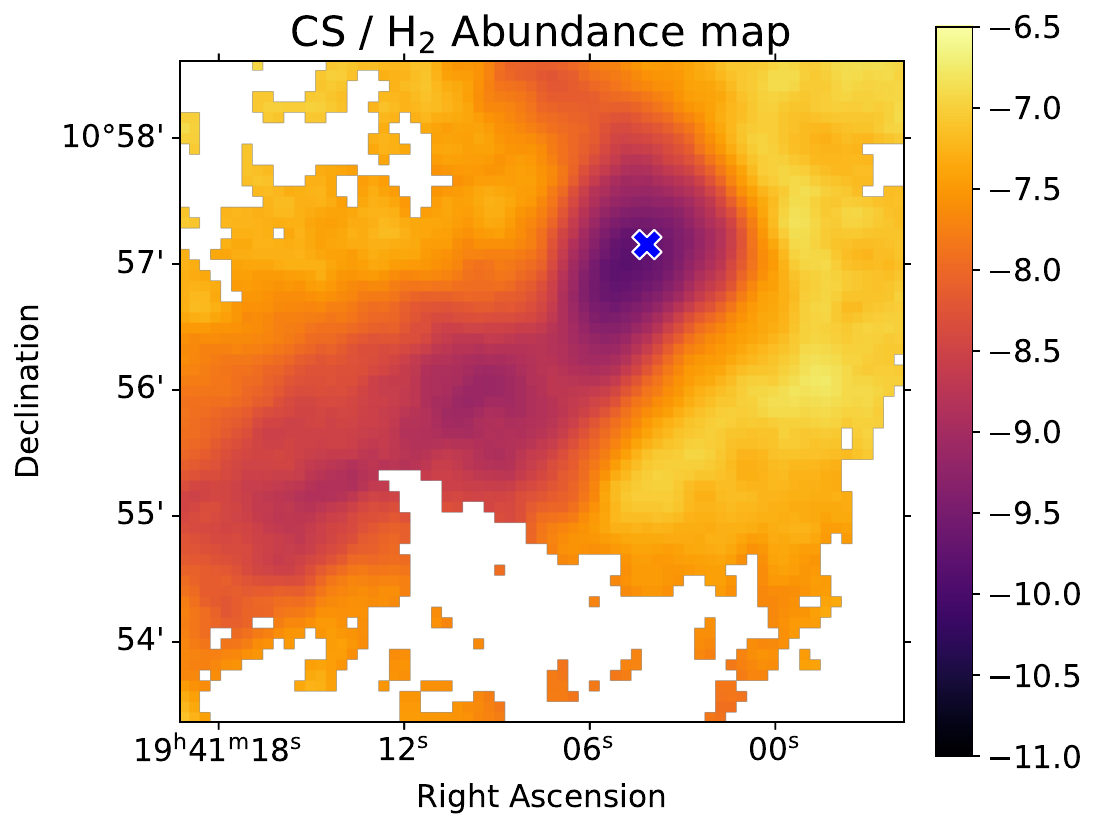}
    \includegraphics[width=0.445\linewidth]{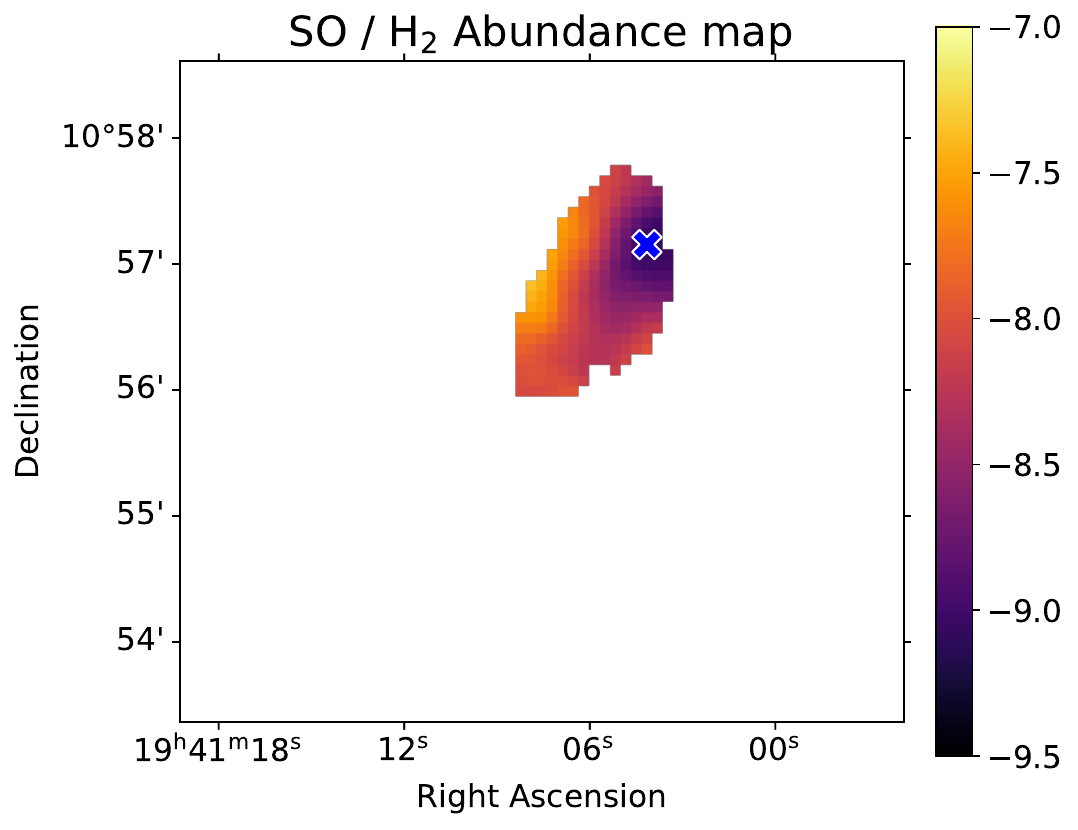}     
    \caption{Gas-phase abundances of observed molecules with respect to H$_2$ on a logarithm scale (log(X$_{mol}$)). The green cross marked on the CH$_3$OH map indicates the position of the background star towards which methanol ice was observed by \citet{chu_observations_2020}. The blue star represents the continuum peak position. }
    \label{fig:l694_abundance}
\end{figure*}

\begin{figure} 
    \centering
    \includegraphics[width=0.9\linewidth]{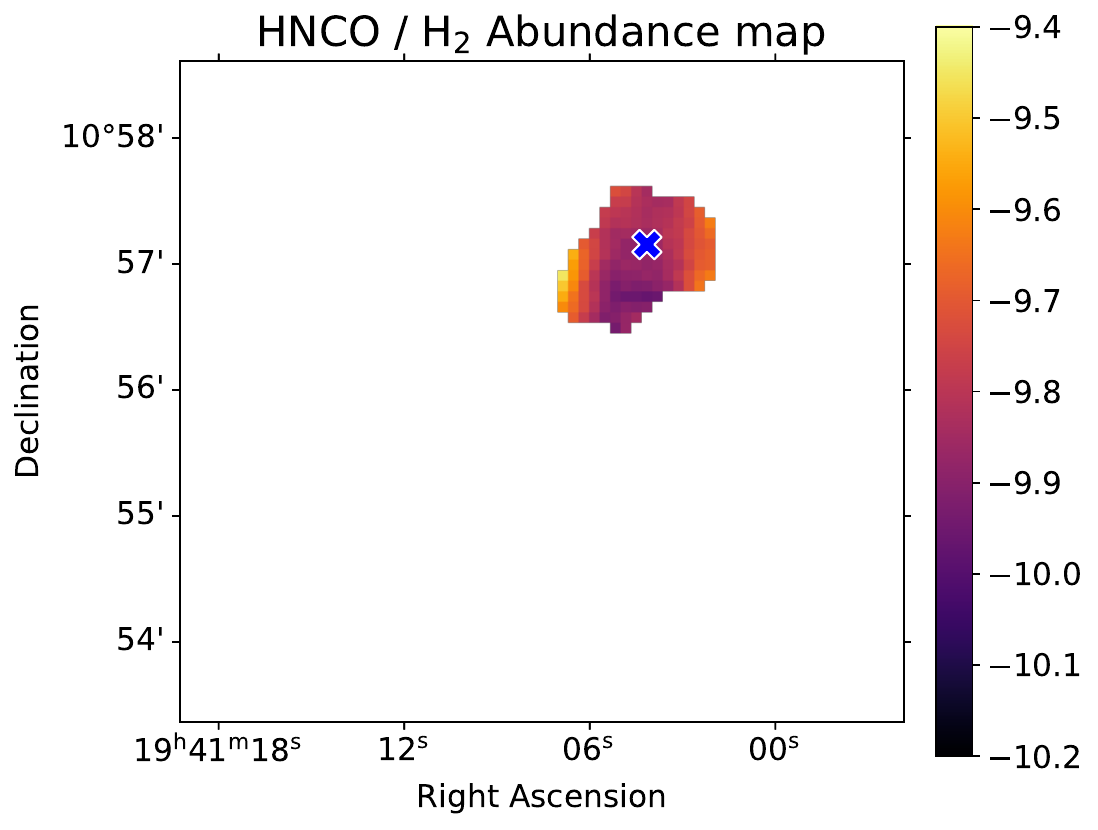}
    \includegraphics[width=0.9\linewidth]{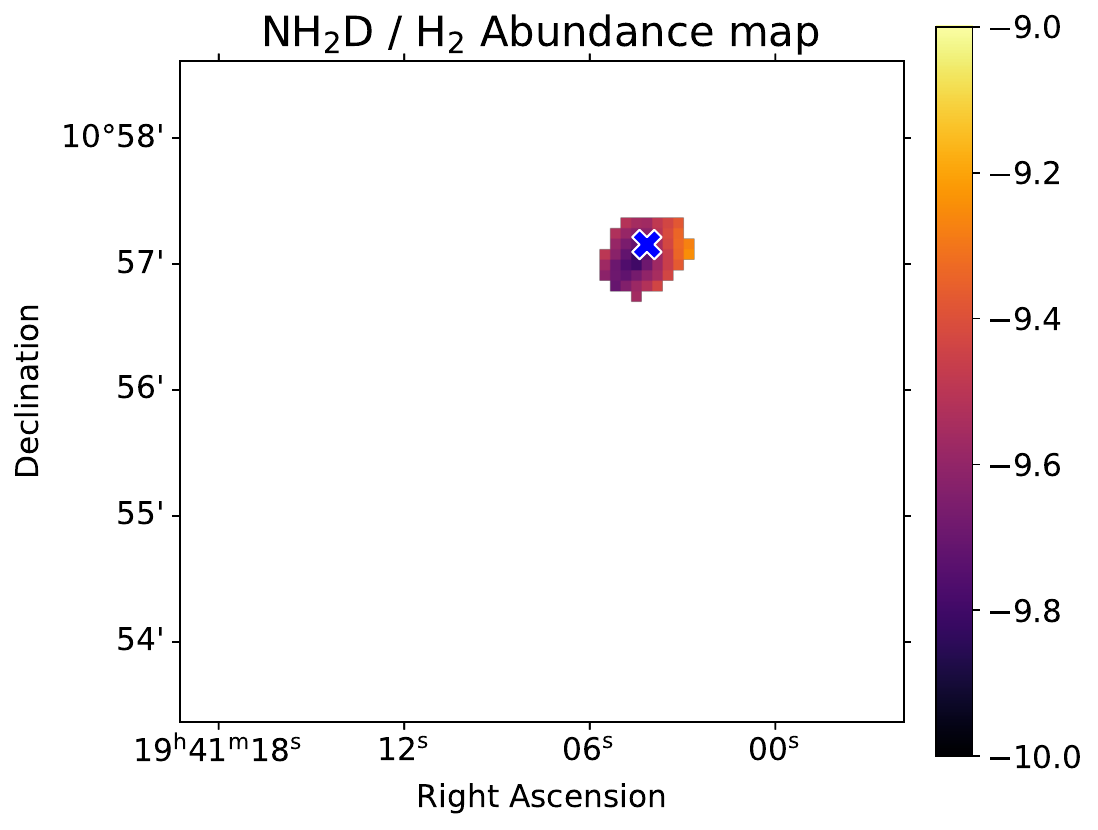}
    \caption{Maps similar to those presented in Fig.~\ref{fig:l694_abundance} for HNCO and NH$_2$D.}
    \label{fig:l694_abundance2}
\end{figure}

The maps obtained using method one are presented in Figs.~\ref{fig:l694_abundance} and \ref{fig:l694_abundance2}.
As can be seen from the decreasing abundances around the blue cross, all the molecules present a depletion profile at the continuum peak.

The CO shows a strong depletion profile at the continuum peak, with its lowest abundance of 3.9 $\times 10^{-5}$ and its maximum of 1.9 $\times 10^{-4}$.
The depletion between the two density extremes observed here is approximately a factor of 7. 
We can compute its depletion factor, f$_{\rm CO}$, by comparison to its canonical abundance measured in molecular clouds. The depletion factor is computed as f$_{\rm CO}$ = f(X$_{\rm can}$/X$_{\rm ^{12}CO}$), where X$_{\rm can}$ = 8.5 x 10$^{-5}$ is the canonical abundance of $^{12}$CO measured by \citet{frerking_relationship_1982} in Taurus and $\rho$ Oph.
We obtain f$_{\rm CO}$ = 2.23. This value is lower than that obtained by \citet{crapsi_probing_2005}, who used the same canonical abundance to find f$_{\rm CO}$ = 11 at the core position. 
The difference between the results mainly arises from the two methods adopted: \citet{crapsi_probing_2005} used Eq. 1 from \citet{caselli2002}, which depends on the integrated intensity of C$^{18}$O and the observed brightness at 1.2 mm. They note that the depletion factor is subject to multiple uncertainties, mainly due to the determination of the canonical abundance and the amount of the molecule actually probed in the densest region. CS exhibits optically thick emission lines from our RADEX computation, meaning that the abundance we determined is an upper-limit on the actual quantity. CS likewise exhibits a significant depletion gradient along the filament, where the abundance decreases to $1.6 \times 10^{-10}$ at the continuum peak. The maximum value, located west of the source, is $1.7 \times 10^{-7}$: three orders of magnitude higher than the minimum value.
For CH$_3$OH, the abundance is highest east of the continuum peak (in a similar position to SO and HNCO), with a value of 9.8 $\times 10^{-9}$, and reaches a minimum of $5.2 \times 10^{-10}$ at the continuum peak. The observational constraint on the ice abundance by \citet{chu_constraining_2021} corresponds to a volume density of $\sim 4.3 \times 10^4$ cm$^{-3}$. Comparing this methanol ice column density with our gas-phase one at the same position, we find a gas-to-ice ratio of approximately 0.003\%. In L429-C, at a similar density of $\sim 2.4 \times 10^4$ cm$^{-3}$, the ratio was approximately 0.002\%. 
N$_2$H$^+$ exhibits hyperfine structure with three distinct peaks. All three transitions trace the same region, with critical densities at 12 K for the transitions spanning a narrow range from 2 $\times$ 10$^4$ to 6.3 $\times$ 10$^5$ cm$^{-3}$. In this case, we computed the column density by fitting only the middle component (with the rest frequency listed in Table~\ref{tab.molecules}), after checking that all hyperfine components were optically thin and traced the same spatial distribution. 
Maximum abundance can be found at the filament border, with a value of 7.7 $\times 10^{-10}$, while minimum abundance, 5.5 $\times 10^{-11}$, occurs at the continuum peak. As the literature \citep{Caselli_N2H_2002,2013A&A...560A..41L} finds, a correlation exists between the CO depletion profile and the presence of N$_2$H$^+$ tracing the dense part of the core. N$_2$H$^+$ is considered a late depleter \citep{Bergin_2002}, and its low abundance at the peak position suggests an aged core. The late depletion of N-bearing species has been confirmed in recent studies \citep{Redaelli_2019_NH2+,Caselli_2022,Lin_2020,Lin_2023_NH3}. In \citet{Pineda_2022}, the authors suggest that the depletion of another N-bearing species, NH$_3$, begins at densities higher than 2 $\times$ 10$^{5}$ cm$^{-3}$. In our case, N$_2$H$^+$ depletion occurs at a lower density ($\sim$ 10$^{4}$ cm$^{-3}$). In \citet{Lin_2020}, the authors find an N$_2$H$^+$ depletion of $\sim$ 27 in the cold core L1512 at a central density of 10$^5$ cm$^{-3}$, while the depletion we observe in L694 is by a factor of $\sim$ 15 at a density of the same order of magnitude.
SO is detected in an even smaller area than N$_2$H$^+$, showing maximum abundance to the east of the source, with a value of $4.8 \times 10^{-8}$. Its minimum abundance is found at the dust continuum peak with a value of 9.7 $\times 10^{-10}$. 
HNCO and NH$_2$D are detected only in a small area immediately surrounding the core. These two molecules exhibit a tighter range of values between their extrema. HNCO abundance at the continuum peak is 1.4 $\times 10^{-10}$, with a maximum found east of this position--similar to SO--at 3.2 $\times 10^{-10}$. For NH$_2$D, the maximum abundance is 5.6 $\times 10^{-10}$, measured next to the continuum peak position, where the abundance is 1.5 $\times 10^{-10}$.

\begin{table}[]
    \centering
    \caption{Maximum abundance and depletion factor.}
    \begin{tabular}{c|c|c}
        \hline
        \hline
        Molecule     & Maximum abundance     & Depletion factor* \\
        \hline
         CO         & 1.9$\times$ 10$^{-4}$ & 7 \\
         CH$_3$OH   & 9.8$\times$ 10$^{-9}$ & 19 \\
         CS         & 1.7$\times$ 10$^{-7}$ & 1062 \\
         SO         & 4.8$\times$ 10$^{-8}$ & 50 \\
         HNCO       & 1.4$\times$ 10$^{-10}$ & 2 \\
         N$_2$H$^+$ & 7.7$\times$ 10$^{-10}$ & 14 \\
         N$_2$D     & 5.6$\times$ 10$^{-10}$ & 4 \\
        \hline
    \end{tabular}
    \tablefoot{*Depletion factor is the maximum abundance over the lowest abundance found at the dust continuum peak.}
    \label{tab:depletion_factor}
\end{table}

\subsection{Molecular abundance as a function of physical parameters}\label{L694_ab_parameters}

\begin{figure*}
    \centering
    \includegraphics[width=0.95\linewidth]{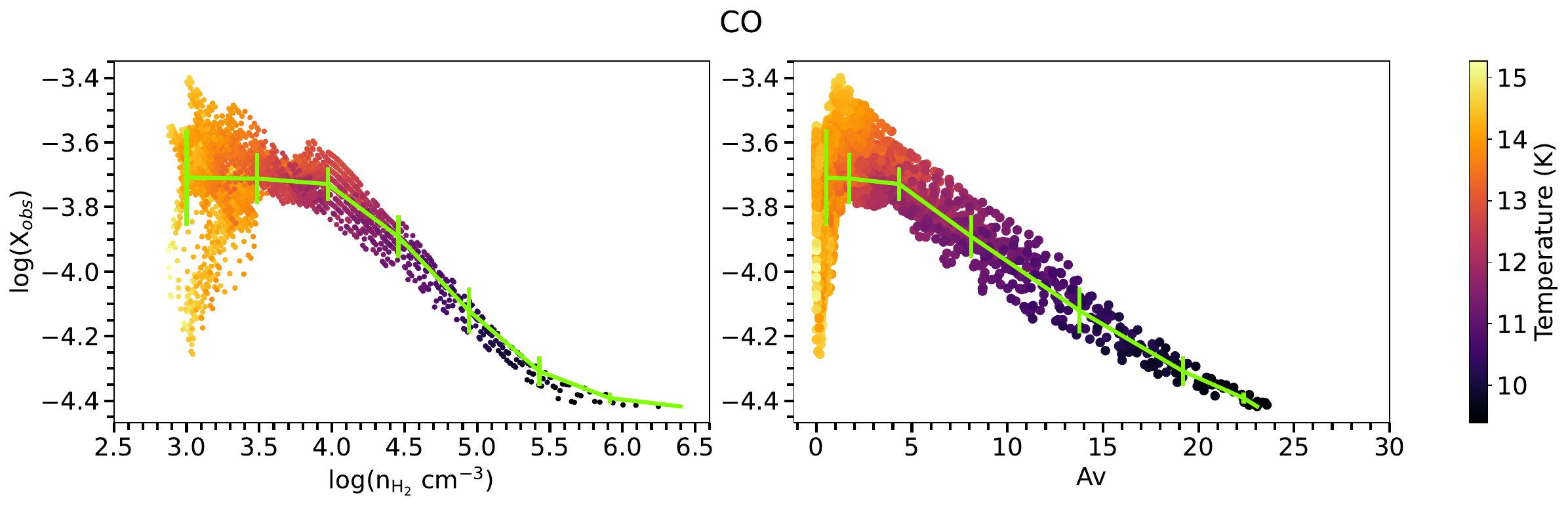}
    \includegraphics[width=0.95\linewidth]{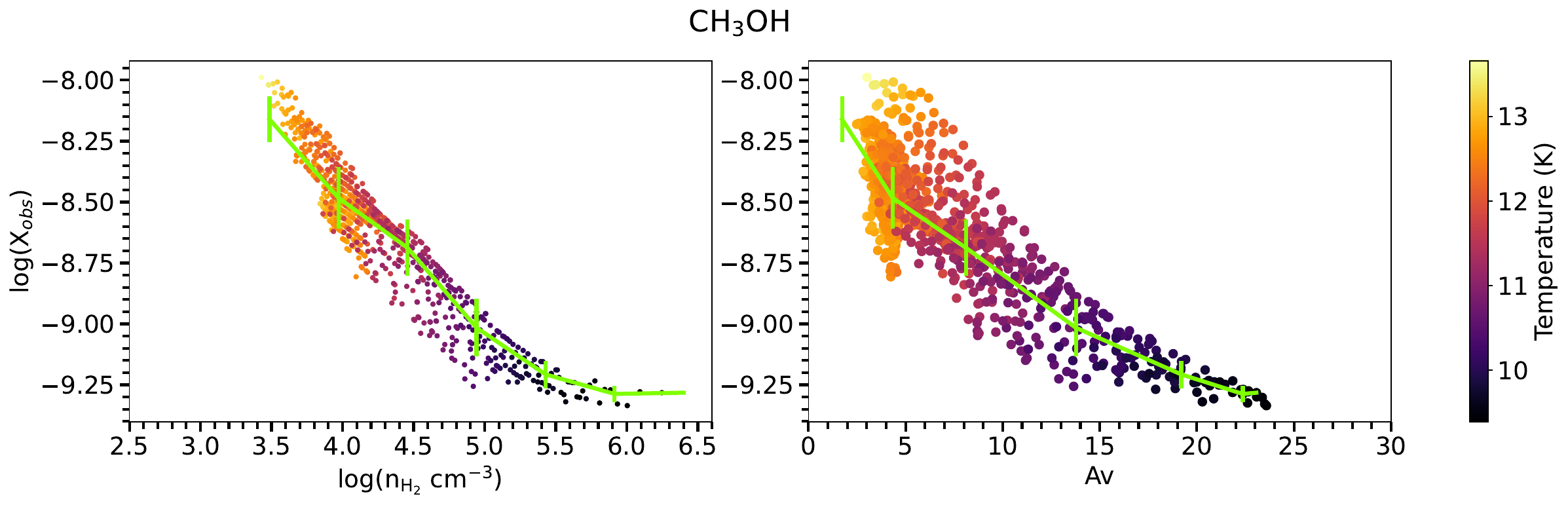}
    \includegraphics[width=0.95\linewidth]{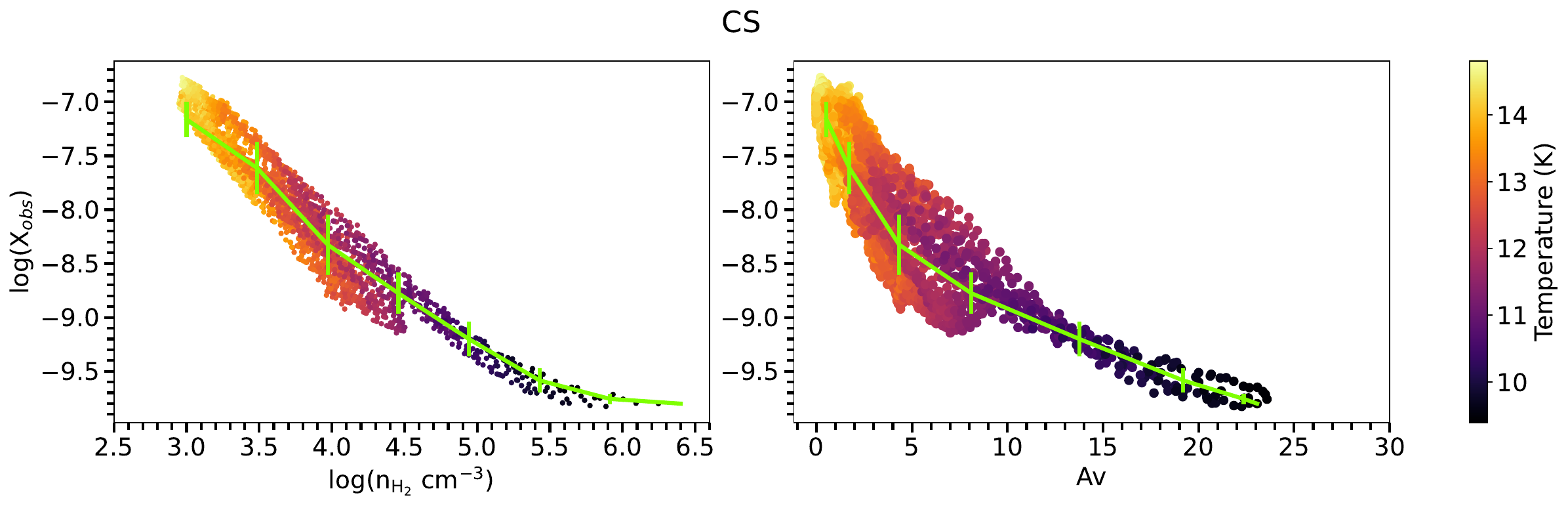}
    \includegraphics[width=0.95\linewidth]{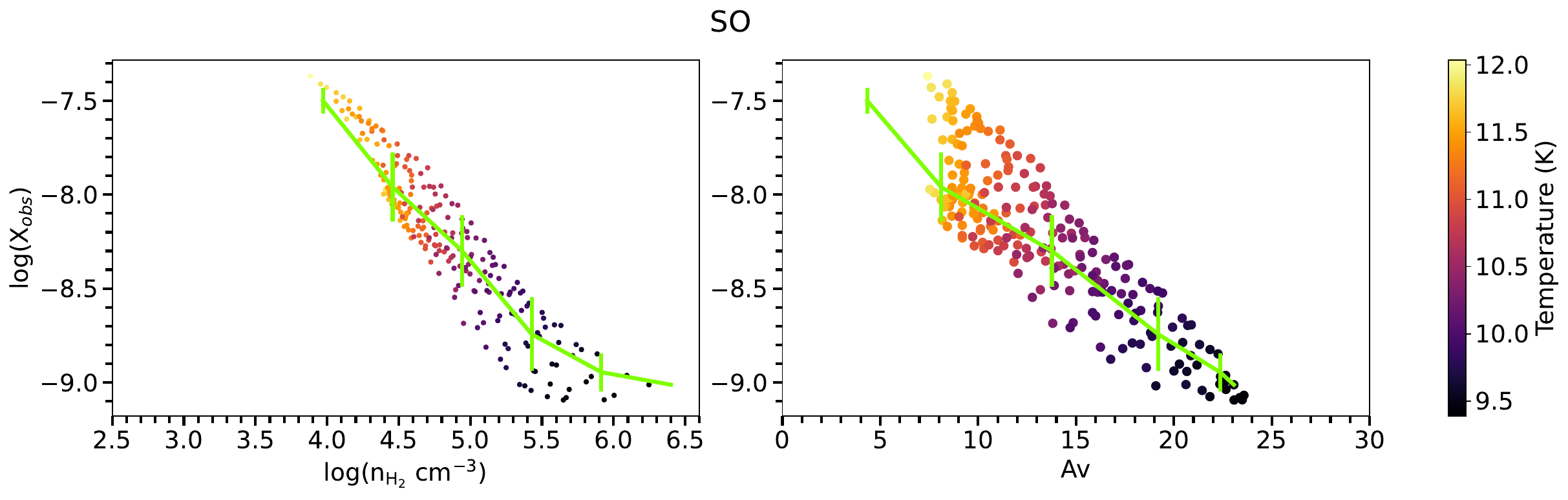}
    \caption{Left: Abundance as a function of hydrogen volume density n$\rm_{H_2}$ for CO, CH$_3$OH, CS, and SO. Right: Abundance as a function of visual extinction, Av. Colour coding indicates the temperature. The mean abundance and standard deviation for each bin appear in green.}
    \label{fig:L694_density_av_vs_abundance1} 
\end{figure*}

\begin{figure*}[h!]
    \centering
    \includegraphics[width=0.95\linewidth]{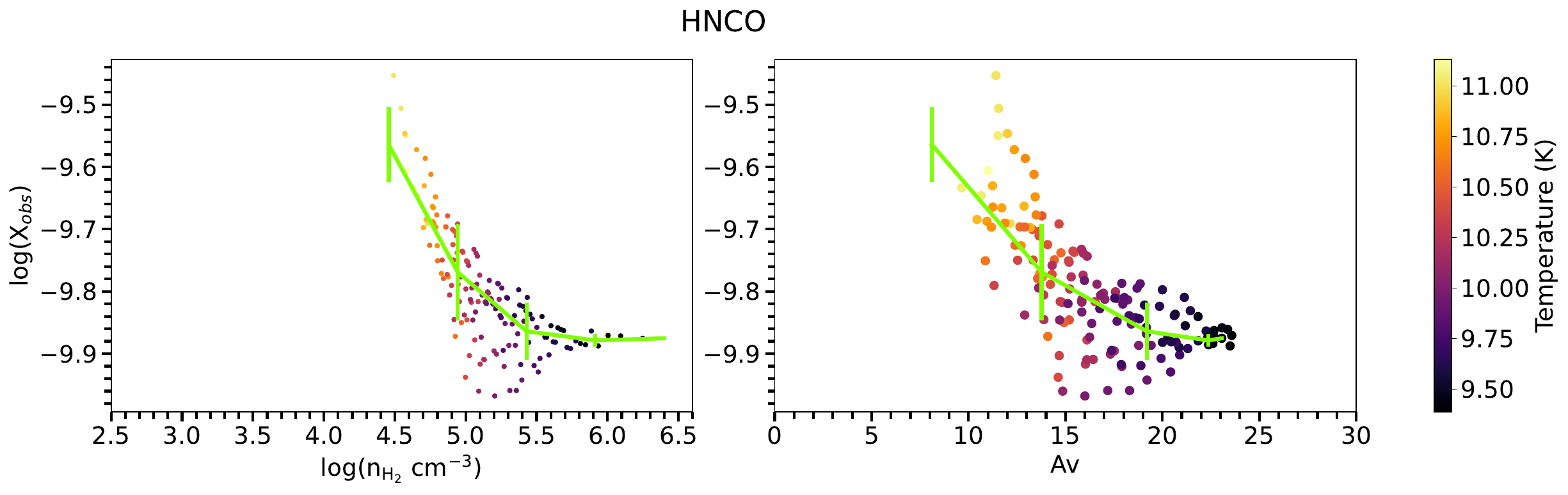} \includegraphics[width=0.95\linewidth]{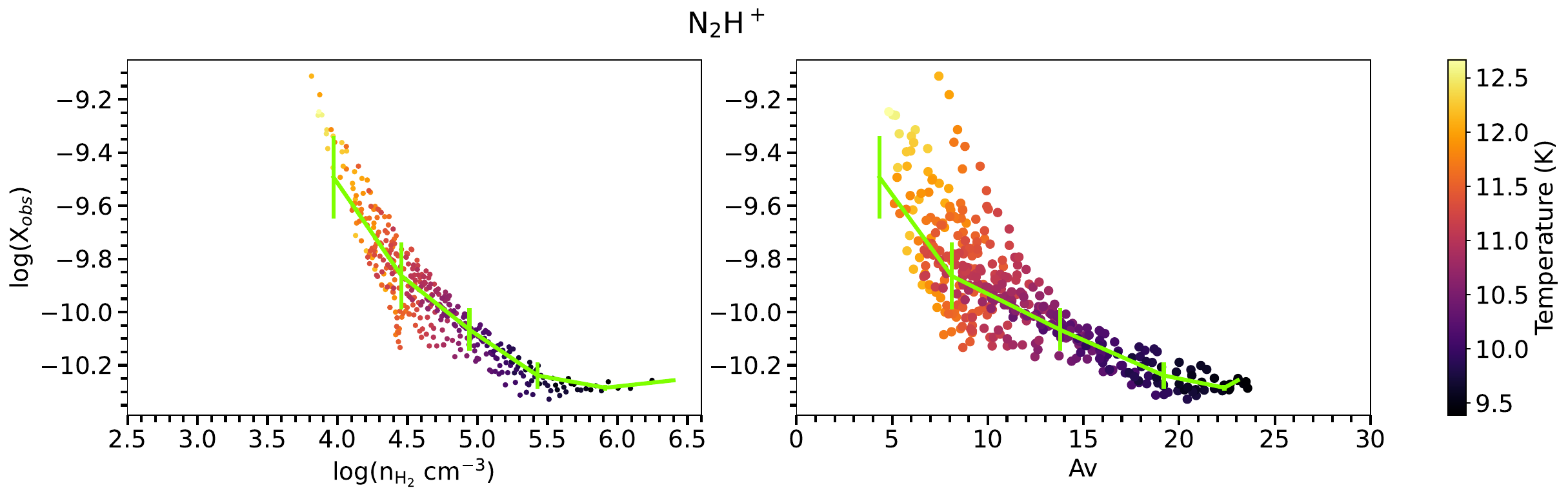}
    \includegraphics[width=0.95\linewidth]{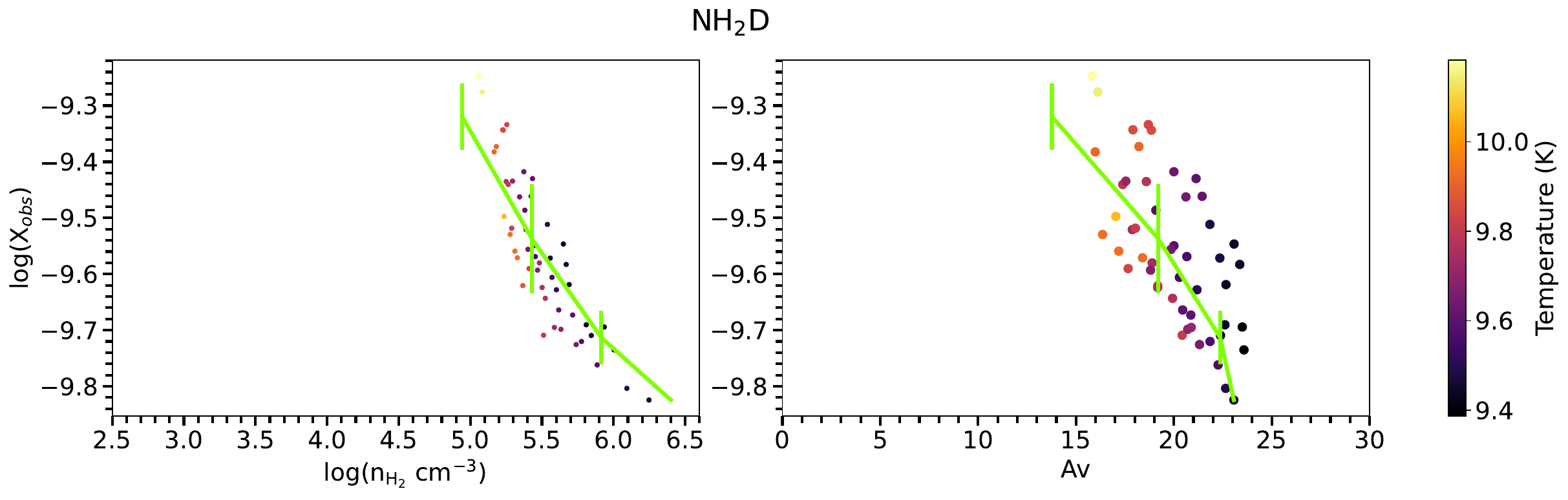}
    \caption{Same as Fig.~\ref{fig:L694_density_av_vs_abundance1}, for HNCO, N$_2$H$^+$, and NH$_2$D.}
    \label{fig:L694_density_av_vs_abundance2}
\end{figure*}

When comparing these molecular abundances (and their standard deviation) to the physical parameters in Figs.~\ref{fig:L694_density_av_vs_abundance1} and \ref{fig:L694_density_av_vs_abundance2}, all molecular abundances decrease with both volume density and visual extinction. We summarise maximum abundances and depletion factors (i.e. the maximum abundance over the abundance at the continuum peak position) in Table~\ref{tab:depletion_factor}.
CO shows a depletion gradient of more than a factor of 5 between volume densities of 2 $\times$ $10^3$~cm$^{-3}$ and 2 $\times$ $10^6$~cm$^{-3}$. 
Its abundance at a lower density seems to plateau where the abundance saturates to a value close to 3.1 $\times$ 10$^{-4}$.
CS has the strongest depletion profile along the filament, with an abundance varying from $1.7 \times 10^{-7}$ to $\sim$ 1.6 $\times 10^{-10}$, a factor $>$ 1000.
CH$_3$OH shows a strong depletion gradient varying from 9.8 $\times$ 10$^{-9}$ to $5.2 \times 10^{-10}$ (by a factor of $\sim$ 20) at densities increasing from 3 $\times 10^3$ to $10^6$ cm$^{-3}$.
SO shows a similar profile to CH$_3$OH, with strong depletion between 10$^4$ and 10$^6$ cm$^{-3}$, as its abundance drops from $4.8 \times 10^{-8}$ to 9.7 $\times$ 10$^{-10}$, corresponding to a decrease by a factor of approximately 50.
Despite their detection in relatively small regions of the map, HNCO, N$_2$H$^+$, and NH$_2$D exhibit a depletion profile with factors of $\sim$ 2, 14 and 4, respectively. 

\section{Comparison with chemical models}\label{section-models}

To compare with our observational results, we ran two sets of simulations using the Nautilus gas-grain model (see Sect.~\ref{nautilus-section}). As we explain in Sect.~\ref{static-models}, we first adopted the fixed physical conditions observed in our region, as used in the radiative transfer analysis. Next, (see Sect.~\ref{dynamic-models}), we used time-dependent physical parameters as computed by a hydrodynamical model to follow the formation of the cores. 
For both sets of simulations, we compared the model results to the observed abundances over the density range where all the molecules were detected (from 3 $\times$ 10$^4$ to 10$^{6}$ cm$^{-3}$).

\subsection{The Nautilus gas-grain model}\label{nautilus-section}

To run both sets of simulations, we used the gas-grain chemical code Nautilus \citep{ruaud_gas_2016,wakelam2024}. The code is available through a github repository\footnote{https://astrochem-tools.org/codes/}. 
Nautilus is a three-phase model that uses the rate-equation approach to compute, at each time step, the abundance of hundreds of chemical species in the solid (i.e. mantle and surface) and gas phase. The gas-phase reaction rates were computed using the latest kida.uva.2024 chemical network \citep{wakelam2024}. The surface reaction network was based on the rate equation method \citep[as described in][]{1992ApJS...82..167H,ruaud_gas_2016,wakelam2024}. The mobilities of the species at the surface of the grains and in the mantle were set as a fraction of the binding energies (0.8 in the mantle and 0.4 at the surface). As a result, diffusion was more efficient on the surface than in the mantle. As photodissociation (both by direct and secondary UV photons) can occur both at the surface and in the mantle. Mechanisms such as whole grain cosmic-ray heating, cosmic-ray sputtering, photo-desorption, and chemical desorption were included in addition to thermal desorption. All details on the processes included in the model are given in \citet{ruaud_gas_2016} and \citet{wakelam2024}.

In our static and dynamical simulations, we used a set of standard parameters described below. We used the initial chemical composition listed in Table 1 of \citet{wakelam_chemical_2021}, except for the sulphur elemental abundance, which was set to 8 $\times$ 10$^{-8}$ to better reproduce the S-bearing molecules, \citep[following the low-metal abundances from][]{Graedel1982}. 
For the cosmic-ray ionisation rate ($\zeta$), we used the prescription by \citet{wakelam_chemical_2021} (i.e. Eqs. 6 and 7), which depends on visual extinction. As in the observational analysis and as described in Sect.~\ref{sec:physical_param}, we assumed that the gas temperature is equal to that determined using method one. 
As in previous studies \citep{taillard_constraints_2023,clement2023}, we did not use this temperature for the dust, since these maps trace the outer region of the cloud where smaller and warmer grains reach temperatures higher by a few Kelvin. A slightly higher grain temperature can strongly affect chemical abundances, especially in the ices.
We therefore adopted the parametric expression for dust temperature derived by \citet{hocuk_parameterizing_2017}, in which the grain temperature varies as a function of Av. Following \citet{clement2023}, we added 1 K to the dust temperature obtained from this method.

\subsection{Static models}\label{static-models}
\subsubsection{Physical parameters}

Using the three-phase model described above, we first ran the model for a grid of physical parameters similar to those obtained with method one (and used to analyse the observed lines), considering only the physical regions where CH$_3$OH is detected. 

We ran two sets of eight models. Each of the eight were sampled on the density range (i.e. from 3 $\times$ 10$^4$ to 10$^{6}$ cm$^{-3}$) previously computed with method one. Visual extinction and gas temperature were computed from these different density samples
 (T$\rm_{gas}$ decreasing from 11.3 to 9.4 K, and Av increasing from 8.3 to 29.8 mag).
These physical parameters were fixed during the entire integration time.
The sputtering yield was measured in experiments as a function of the predominant molecule in ices, showing that it decreases in H$_2$O-rich ices, while CO$_2$-rich ices exhibit the opposite behaviour \citep{dartois_cosmic_2018,dartois_non-thermal_2020}. This parameter is particularly important for the non-thermal desorption of certain molecules, especially methanol \citep{wakelam_efficiency_2021,taillard_constraints_2023}. 
For each set of eight models, we used either the H$_2$O-rich or CO$_2$-rich sputtering yield. In total, there are 16 unique models, which we summarise in Table~\ref{tab:models}.
We then ran the models for each of the physical conditions, starting from the initial conditions described in Sect.~\ref{nautilus-section}, over a period of $10^7$ yr.

\begin{table}[]
    \centering
    \caption{Physical parameters of the static models.}
    \begin{tabular}{c|c|c|c}
    \hline
    \hline
     Model        & log(Density) (cm$^{-3}$) & T$\rm_{gas}$ (K) & Av (mag) \\
    \hline
     1 \& 9*          & 4.5              & 11.3         & 8.3   \\
     2 \& 10          & 4.7              & 10.8         & 11.1  \\
     3 \& 11           & 4.9              & 10.4         & 14.8  \\
     4 \& 12           & 5.1              & 10.0         & 19.1  \\
     5 \& 13           & 5.3              & 9.8          & 23.3  \\
     6 \& 14           & 5.6              & 9.6          & 26.7  \\
     7 \& 15           & 5.8              & 9.5          & 28.7  \\
     8 \& 16           & 6.0              & 9.4          & 29.8  \\
    \hline
    \end{tabular}
    \tablefoot{*For each set of physical conditions (n(H$_2$), T, Av), we ran two models using either the H$_2$O-rich or the CO$_2$-rich CR sputtering yield, see text.}
    \label{tab:models}
\end{table}

\subsubsection{Results}\label{staticmodels-results}

\begin{figure}[h!]
    \centering
    \includegraphics[width=0.99\linewidth]{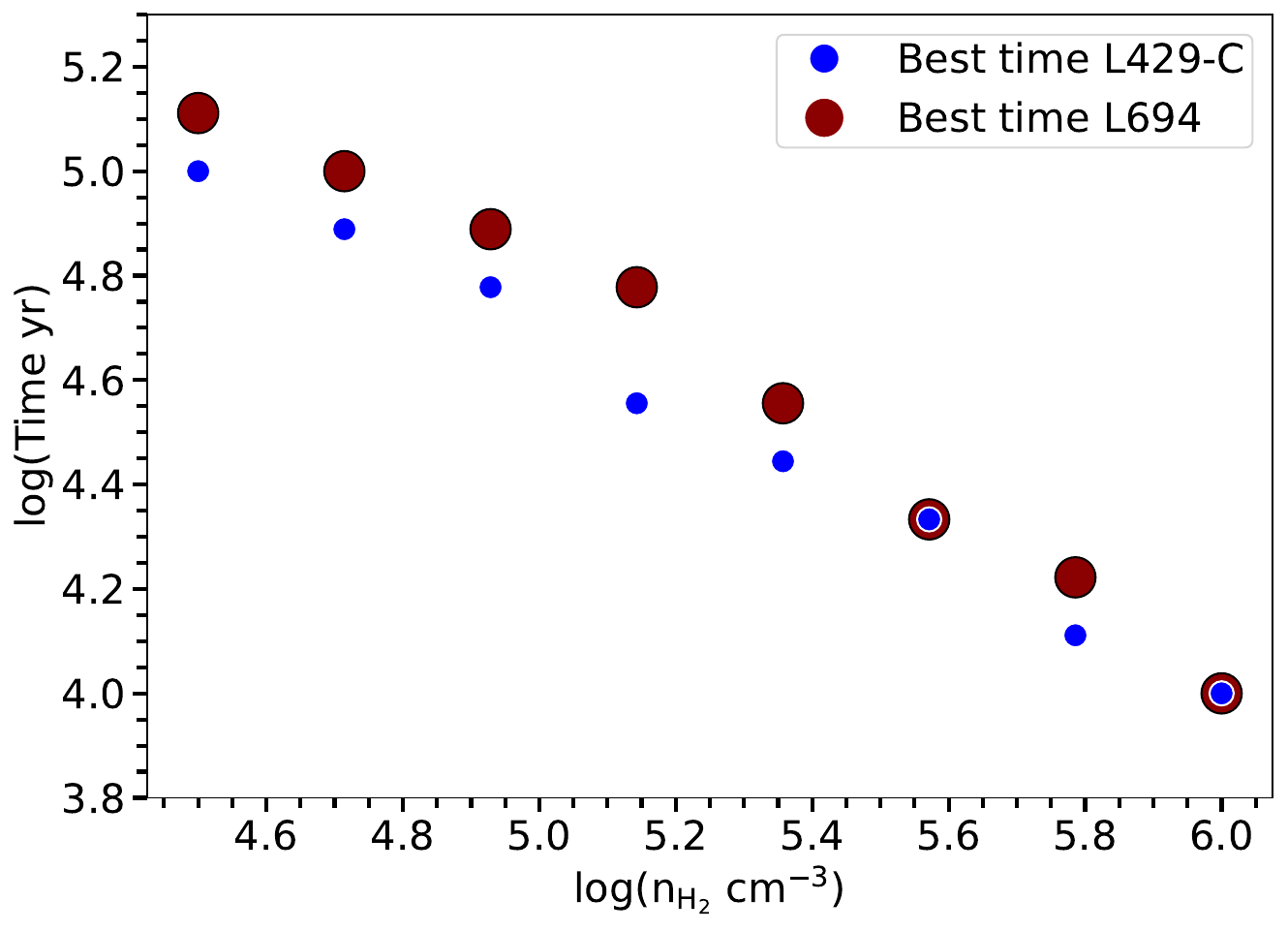}
     \caption{Best time from the static models, obtained using the distance of disagreement based on three observed molecules (CO, CS, and CH$_3$OH), as a function of density (logarithmic scale). Brown circles correspond to L694, and blue circles to L429-C, as obtained from our previous study \citep{taillard_constraints_2023}.}
    \label{fig:L694_best_time}
\end{figure}

To select the time that best reproduces our observations for each physical condition, we used the distance of disagreement--as described in \citet{wakelam_effect_2006}--applied to three observed molecules (CS, CO, and CH$_3$OH), which are detected over the largest range of density in the cloud. 
The distance of disagreement defines the 'best time' as the chemical integration time at which the modelled abundances are the closest to the ones observed. The observed abundances used for comparison with the models are the average values within each density bin, where the observed density was binned according to the density parameters adopted in the models (see Table~\ref{tab:models}).
The best time as a function of density is shown in Fig.~\ref{fig:L694_best_time}, where we observe a general trend across all the sets of models: the time required to reproduce the observations decreases with density, from 1.3 $\times 10^5$ yr at 3.1 $\times$ 10$^4$ cm$^{-3}$ to 1.0 $\times 10^4$ yr at 1.0 $ \times$ 10$^6$ cm$^{-3}$. The two sets of models yield similar results. As in \citet{taillard_constraints_2023}, where we applied the same method to another core, the constraint is mainly given by CO, whose abundance was primarily set by the timescale required for depletion onto the grains \citep[see also discussion section 3 in][]{wakelam_chemical_2021}, rather than by the cosmic-ray sputtering yield, which only affects abundance after depletion has occurred. In appendix~\ref{goodnessoffit}, we discuss the level of agreement for the main molecules at the best times.  

\begin{figure}[h!]
    \centering
    \includegraphics[width=0.99\linewidth]{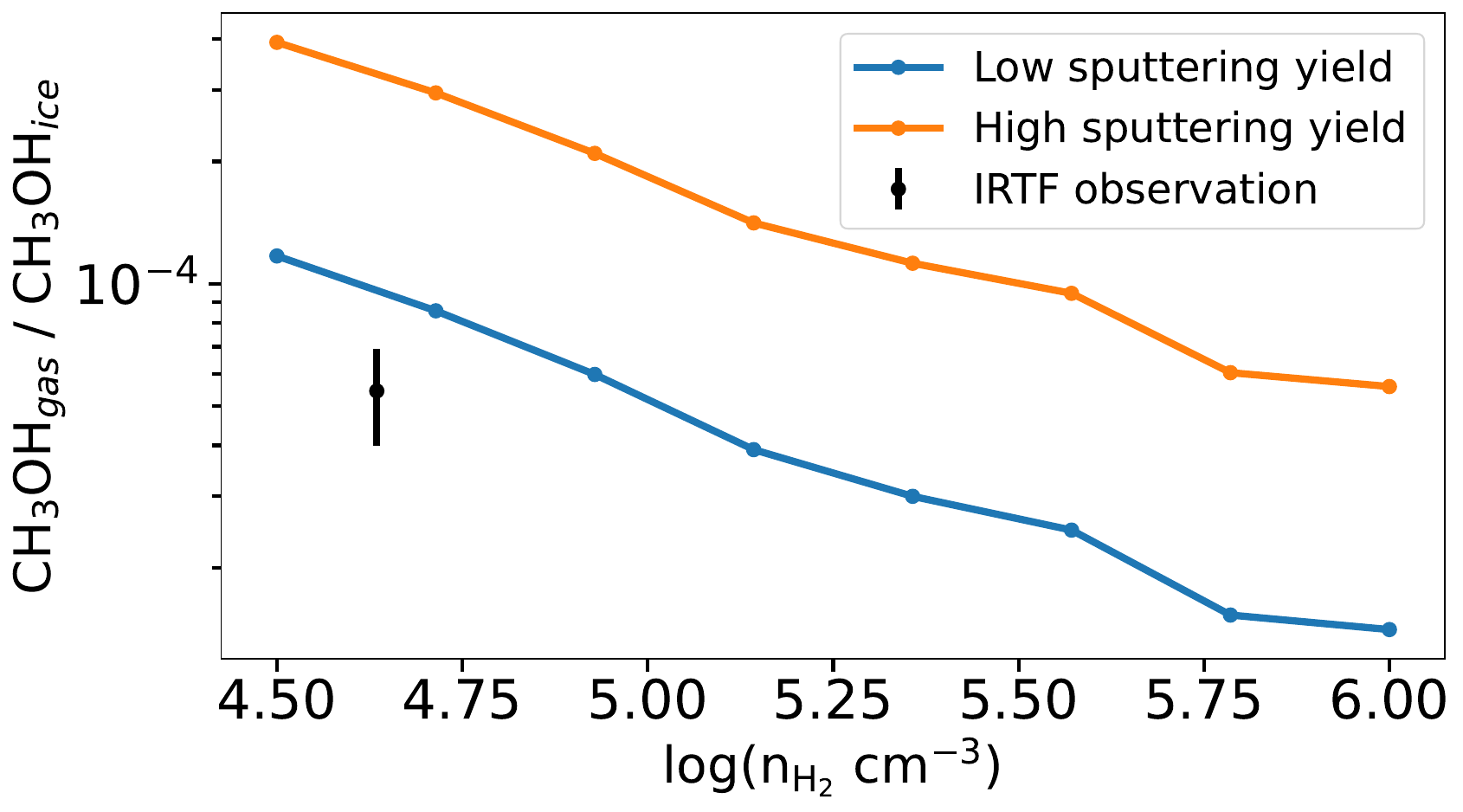}
     \caption{Methanol gas-to-ice ratio obtained for the two different sets of models as a function of density (in logarithm scale). The low and high CR sputtering yield appear in blue and orange, respectively.  In black, we plot the IRTF observation from \citet{chu_observations_2020} and its associated error.}
    \label{fig:L694_CH3OH_ice_to_gas}
\end{figure}

In Fig.~\ref{fig:L694_CH3OH_ice_to_gas}, we compare the gas-to-ice ratio measured in L694 by \citet{chu_observations_2020} with our best models. We find that the model with the low sputtering yield (shown in blue) produces a ratio closer to the observed value (0.008\% at 5.1 $\times 10^4$ cm$^{-3}$, as indicated by the black dot), while the models with the higher sputtering yield (shown in orange) differ from the observations by a factor of 10.
At the density corresponding to the IRTF observations, we also compared our model results with the CO ice abundance (not shown in the figure). We find that all the models reproduce the observed CO ice abundance exactly, at approximately 1.2 $\times$ 10$^{-4}$. 

In Appendix~\ref{annexe-Chu}, we present the results of the static chemical models using the physical parameters derived with method two. We also compare these results with the observed gas-phase abundances.

\subsection{Dynamical models}\label{dynamic-models}
\subsubsection{Physical parameters}

In previous studies, we showed that the timescale evolution between the diffuse medium and the dense core impacts the computed chemical composition relative to the static models described above \citep{2018A&A...611A..96R,2019MNRAS.486.4198W,clement2023}. As suggested by the comparison between the static chemical models and the observations \citep[see also][]{taillard_constraints_2023}, the densest parts of the cold core may experience a faster evolution of their density than the external parts. We therefore compared our observations to chemical simulations in which the physical parameters evolve with time, rather than remain fixed. The time dependent physical parameters were computed with the Smooth Particle Hydrodynamical (SPH) model presented in \citet{2013MNRAS.430.1790B}. These simulations follow the gas dynamics in a galactic potential including spiral arms. In a 250 $\times$ 250 pc$^2$ region of the simulation, we identified 12 cold cores that formed at the end of the simulation (i.e. with a density larger than $10^5$~cm$^{-3}$). We selected all the particles in a region of 0.5 pc in a radius around the peak density and reconstructed the history of each of these particles in terms of density, gas temperature, and visual extinction. The number of particles per core varies between 80 and 350. A full description of these simulations can be found in \citet{2018A&A...611A..96R}. These time-dependent physical conditions were then used as inputs for the Nautilus gas-grain model. The other parameters (i.e. $\zeta$, elemental abundances, and computation of dust temperature) are as described in section~\ref{nautilus-section}. In the next section, we present one core as an example; the conclusions are consistent across all simulated cores.

\subsubsection{Model results}

\begin{figure*}[htbp!]
    \centering
    \includegraphics[width=0.8\linewidth]{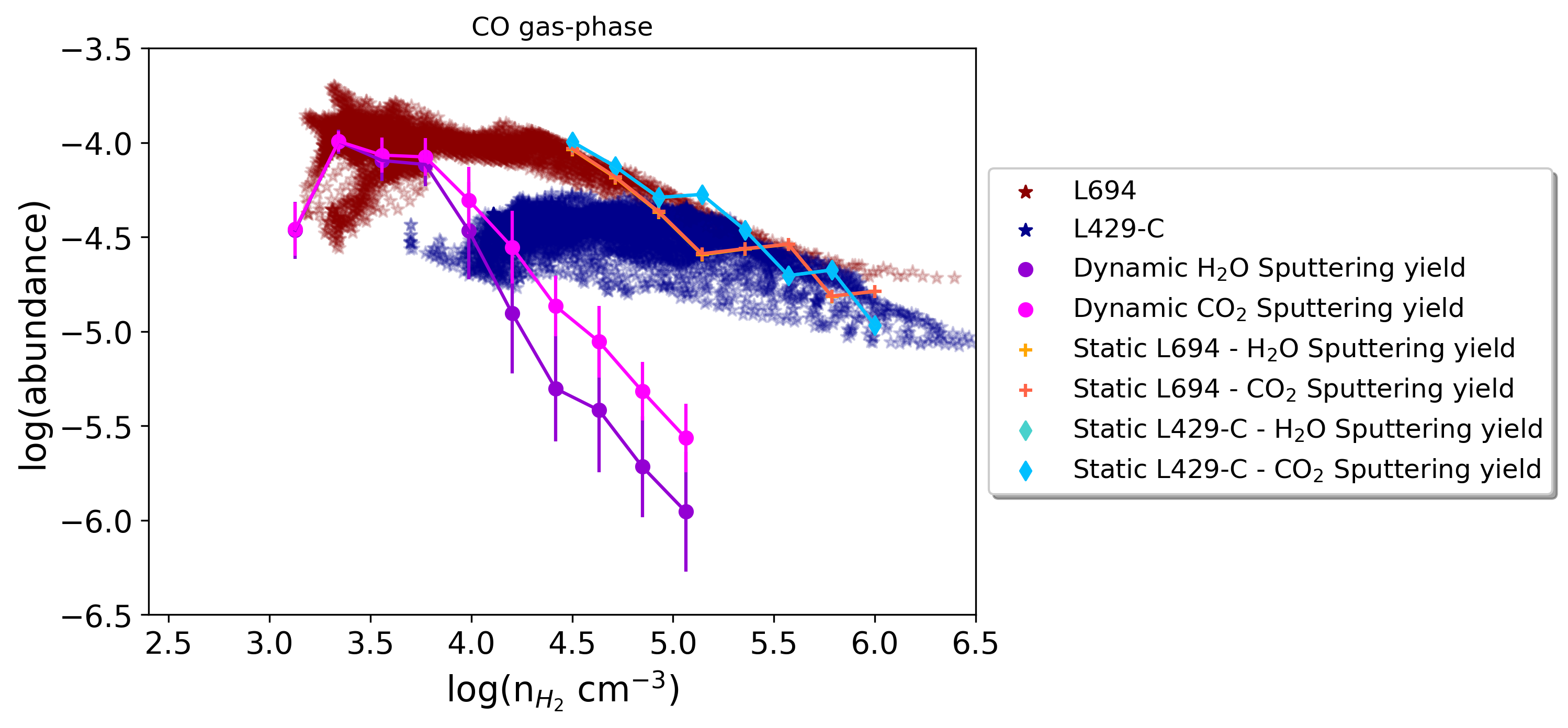}
    \includegraphics[width=0.49\linewidth]{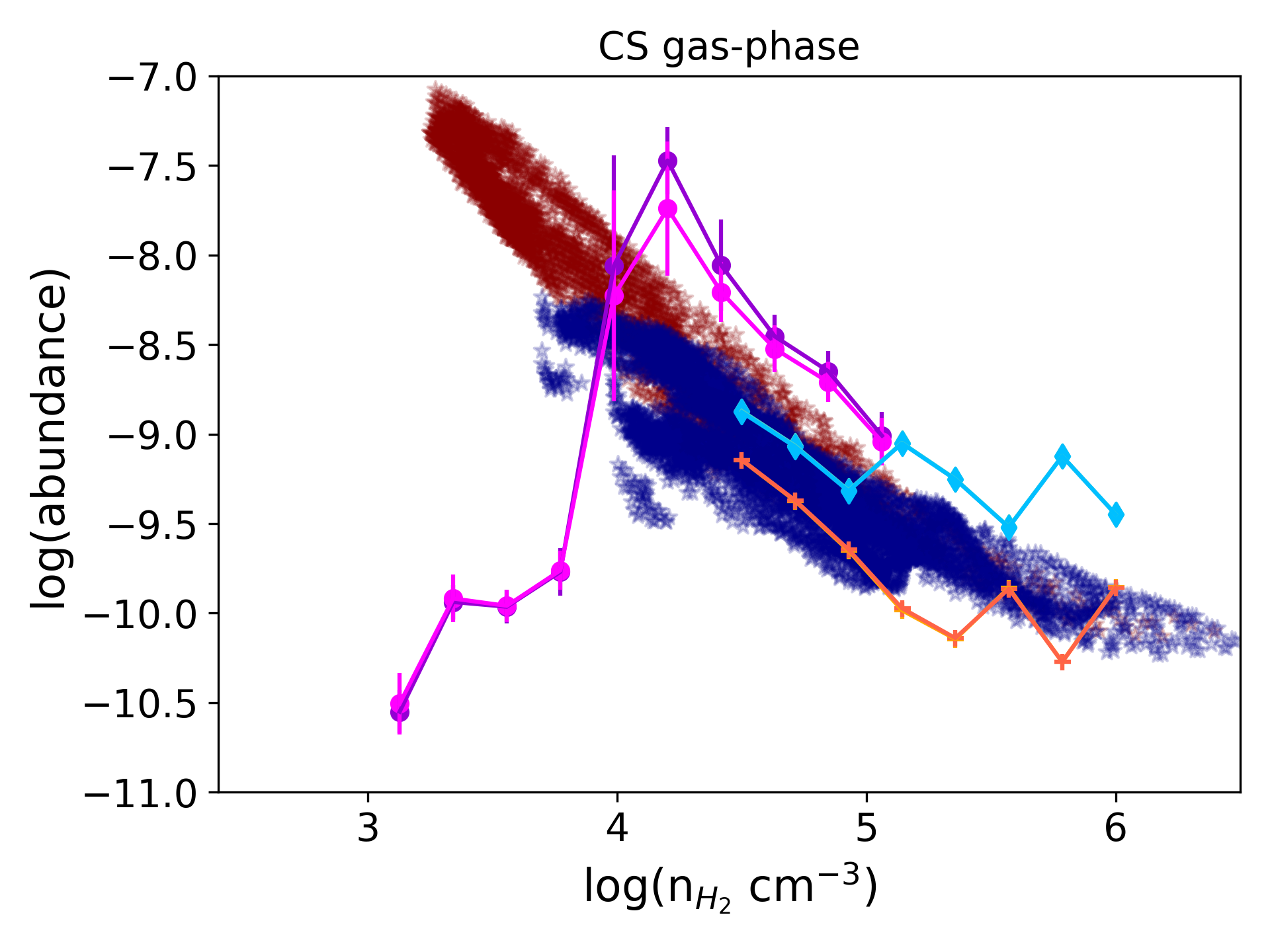}
    \includegraphics[width=0.49\linewidth]{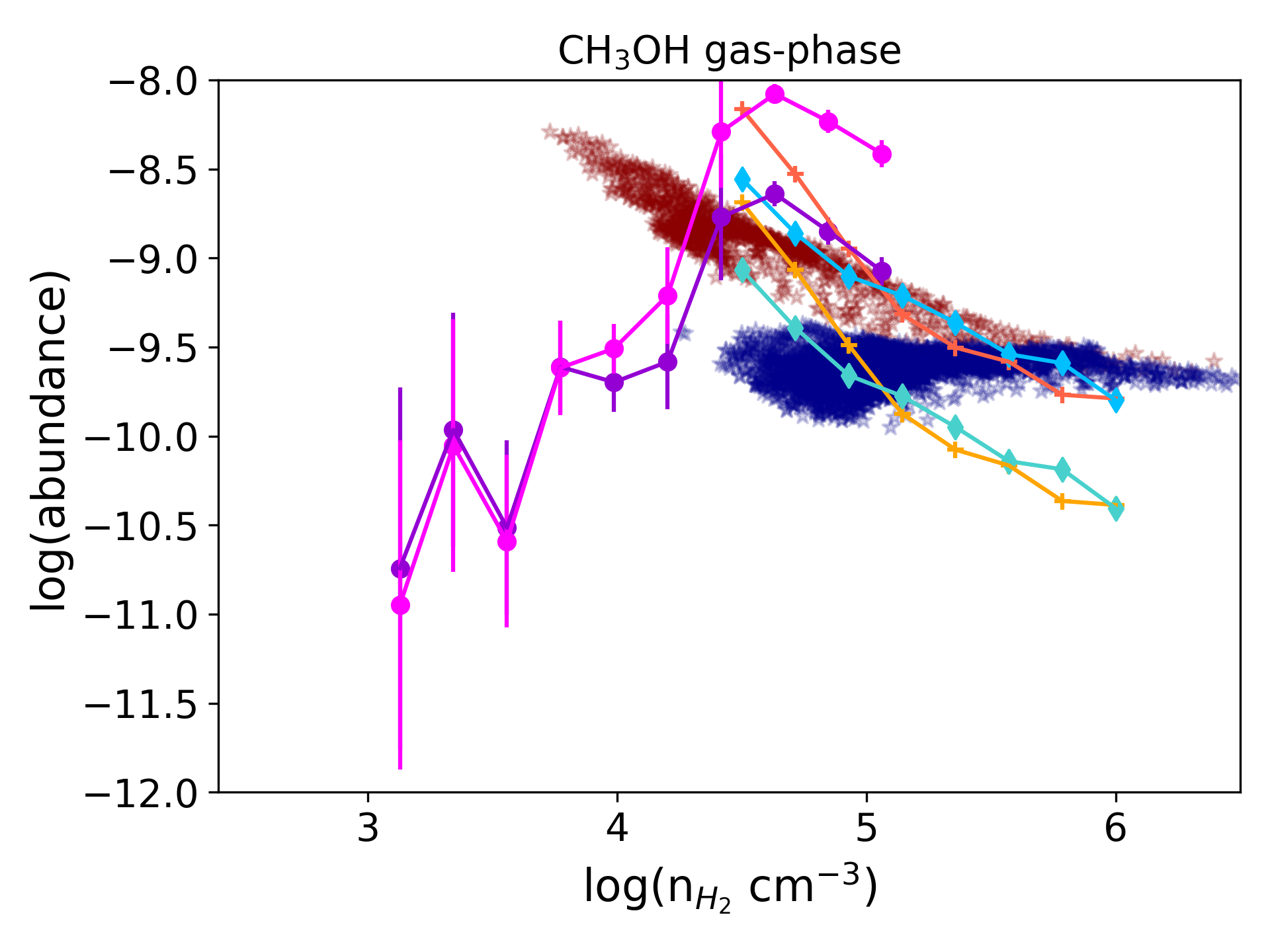}
    \includegraphics[width=0.49\linewidth]{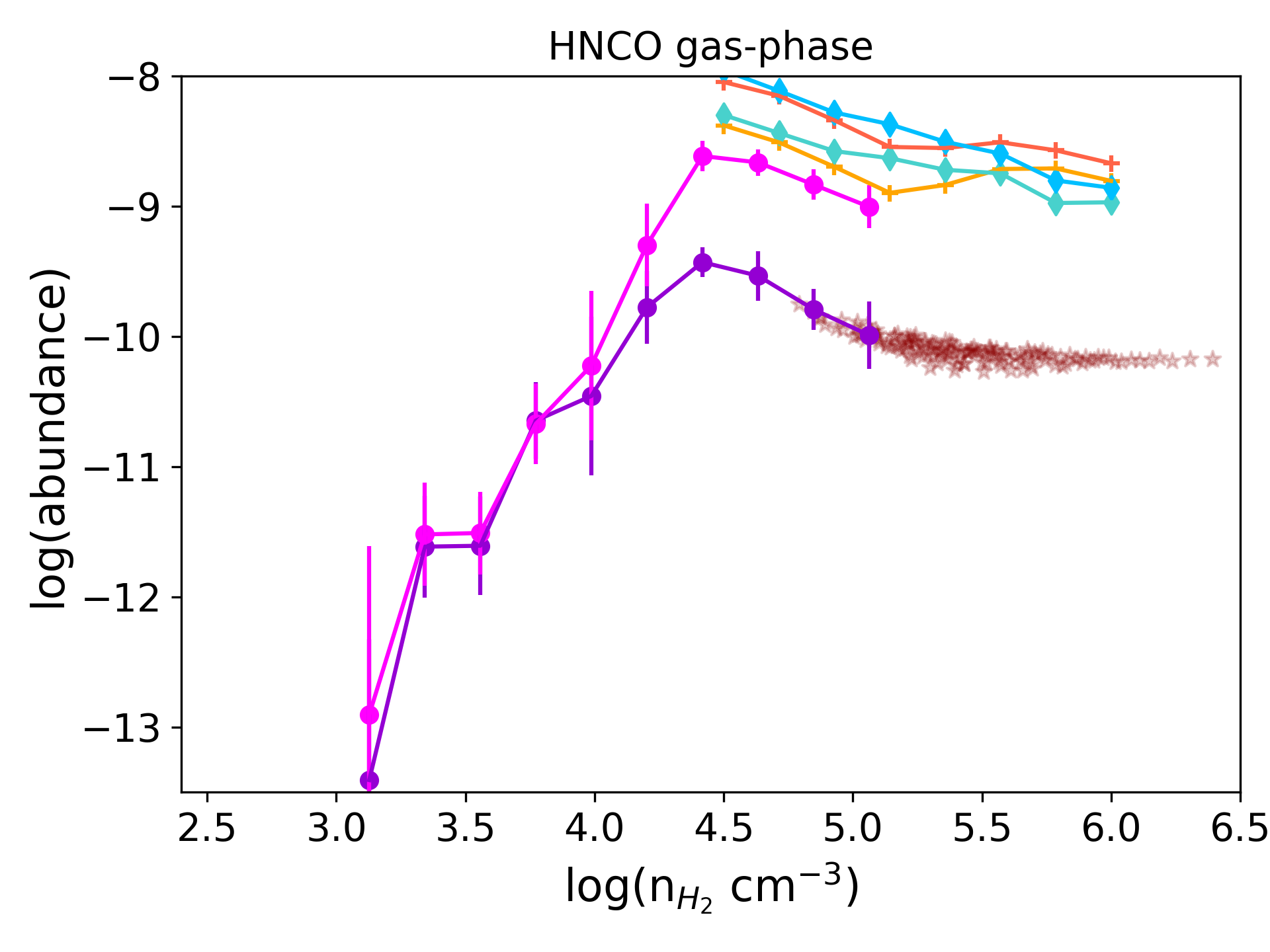}
        \includegraphics[width=0.49\linewidth]{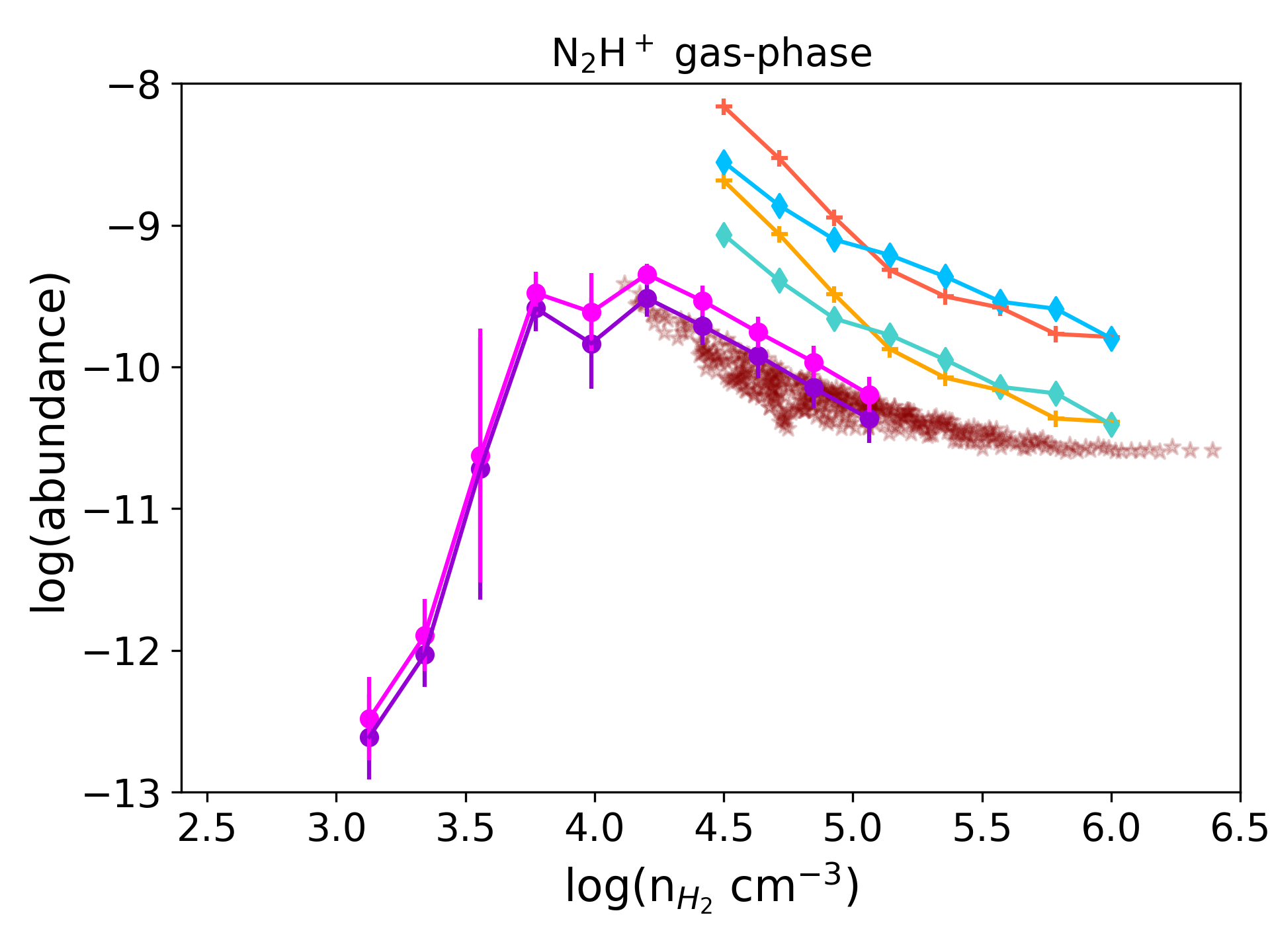}
    
     \caption{Mean abundances computed with the dynamical model (filled circles) as a function of density (see text). Standard deviation for the cold core is shown with vertical bars. Red and blue stars mark the observed abundances in L694 (this study) and L429-C \citep{taillard_constraints_2023}, respectively. Crosses and diamonds denote the abundances obtained at the best times with the static models, which only probe the density where the three molecules (CO, CS, and CH$_3$OH) used to determine the 'best time' are detected. Results obtained from using both cosmic-ray sputtering yields are shown for all the models.
}
    \label{fig:dynamical-simulations}
\end{figure*}

To compare the model results with the observations, we used the output of the chemical composition at the time of core formation. As previously shown \citep{2018A&A...611A..96R,clement2023}, the various histories of the different particles produce a spread in the chemical abundances within each core. As such, for a sampling of the final densities, we computed the mean molecular abundances and standard deviation for each core. In Fig.~\ref{fig:dynamical-simulations}, we show the model results for one selected core as a function of density and superimpose the observed abundances in L694 (this work) and L429-C \citep{taillard_constraints_2023}. We discuss the comparison between these two cold cores in the next section. In the simulations, the abundance of molecules (i.e. CO, CS, CH$_3$OH, HNCO, and N$_2$H$^+$) increase with density before decreasing. For CO, the models reproduce the observations up to a density of a few $10^4$~cm$^{-3}$, but fail at high density. In fact, the modelled CO gas-phase abundances, computed with the H$_2$O-rich cosmic-ray sputtering yield, are lower by more than one order of magnitude at $10^5$~cm$^{-3}$ and above. Using the higher cosmic-ray sputtering yield appropriate for CO$_2$-rich ice, the predicted CO gas-phase abundance at high density increases, but remains significantly lower than that observed. For CS, the models (for both values of the sputtering yield) reproduce the decrease in abundance at densities above a few $10^4$~cm$^{-3}$, but fail to reproduce the abundances at lower densities, where the modelled values are much lower than the observations. Both sets of models nevertheless reproduce N$_2$H$^+$ well. The constraint on the CR sputtering yield is particularly strong for HNCO, which is also well reproduced by the models using the H$_2$O-rich CR sputtering yield. However, since both HNCO and N$_2$H$^+$ are only observed at high density, the model predictions at low density cannot be tested. Methanol is the only molecule for which the agreement with the dynamical simulations is poor across the parameter space. 
Further discussion on the static and dynamical model results is provided in Sect. \ref{discussion-time}.

\section{Discussion}\label{section-discussion}
\subsection{Comparing the L429-C and L694 cold cores}

In our previous study \citep{taillard_constraints_2023}, we applied the same methodology to L429-C, a cold core located in the Aquila Rift, in a less advanced stage (not yet infalling). 
Comparing the physical parameters for both sources using method one, the visual extinction towards the central part of the core is approximately a factor of two higher in L429-C than in L694, with Av $\sim$ 60 in L429-C and $\sim$ 30 in L694. In L694, we also probe regions with lower densities ($10^3$ cm$^{-3}$) compared to L429-C (5 $\times$ $10^3$ cm$^{-3}$). The minimum temperature is similar in the two sources, reaching down to $\sim$ 9 K in L694 and to $\sim$ 10 K in L429-C.
The observed abundances as a function of density in the two sources can be seen in Fig.~\ref{fig:dynamical-simulations}. The CO gas-phase abundance is less depleted in L694 than in L429-C at both low (5 $\times$ $10^3$ cm$^{-3}$) and high density ($10^6$ cm$^{-3}$), and is similar between these densities. The CH$_3$OH abundance is higher in L694 for densities lower than 3.1 $\times$ $10^5$ cm$^{-3}$, while it is similar in the two sources above. Unlike in L429-C, the methanol gas-phase abundance is not flat but decreases with density. The CS gas-phase abundance is similar in both sources. HNCO and N$_2$H$^+$ were observed in L429-C. 
We compare the models' best times in Fig.~\ref{fig:L694_best_time}, which shows the results for the two cores: L429-C (in blue) and L694 (in red). For both pre-stellar cores, the high and low sputtering yield models overlap. The difference between the two sources is minor. Nevertheless, it appears that the timescale of evolution is a bit shorter in L429-C at low density. The external parts of this source might be chemically younger than those of L694. This would be consistent with the fact that this source is at a less evolved stage. We note that, despite the CO abundance difference observed at high density (by a factor of $\sim$ 2), the model is unable to find a different evolutionary time.

\subsection{Using molecular abundances to trace the dynamical evolution of the cores}\label{discussion-time}

\begin{figure}[htbp!]
    \centering
    \includegraphics[width=1\linewidth]{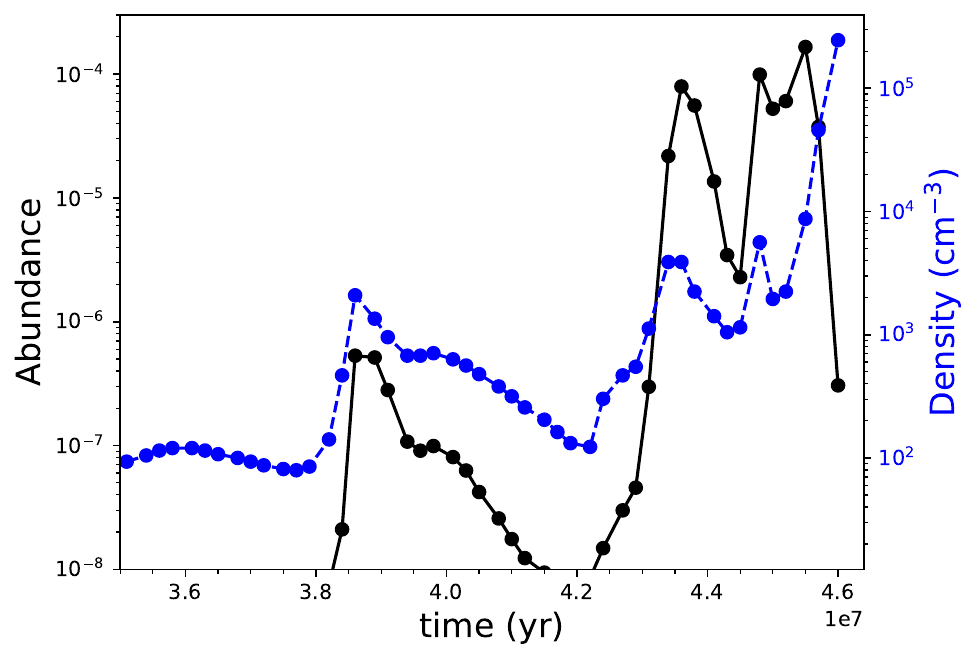}
     \caption{Abundance of CO in the gas phase (black points) and density (blue points) as a function of time computed for one core in the dynamical simulations (see text).}
    \label{fig:trajectory}
\end{figure}

With the static chemical models, we find that the densest parts of L694 (up to densities of $10^6$~cm$^{-3}$) should take approximately $10^4$ years to form from the diffuse medium to reproduce the observed molecular abundances. This constraint was obtained by comparing the observed and modelled molecular abundances, under the assumption that the gas is initially atomic. In the dynamical simulations, the timescale of the density increase is computed by the SPH simulations and thus depends on the physics and the kinematics of the regions. 
If we consider the CO gas-phase abundance, the dynamical model strongly under-produces the observed values at this density (Fig.~\ref{fig:dynamical-simulations}). In these simulations, CO forms early once the density exceeds a few $10^3$~cm$^{-3}$. Figure~\ref{fig:trajectory} shows the results of one trajectory from the dynamical simulations of the selected cloud. In this example, the final drop in the CO gas-phase abundance occurs when the density becomes higher than a few $10^4$~cm$^{-3}$. It takes $5\times 10^5$~years to go from $10^4$ to $2\times 10^5$~cm$^{-3}$. This timescale is much longer than the 'best' time obtained from the static simulations. Carbon monoxide has a robust chemistry, and its gas-phase abundance is a good indicator of the depletion processes at play \citep[see also the discussion in Sect. 3 in][]{wakelam_chemical_2021}. Our results show that the time-dependent density evolution in the dynamical simulations we used is incompatible with the observed high abundance of CO, assuming that CO sticking is correctly reproduced in our model. Three parameters can impact CO depletion: density, time, and the sticking coefficient. Since we have observational constraints on density, time and the sticking coefficient represent good avenues for model improvement. Apart from these physical parameters, both the chemistry and physics of CO can also directly impact its gas-phase abundance. For example, the efficiency of CO conversion into other species \citep[e.g. COMs,][]{fuchs_hydrogenation_2009,Chuang_2016,simons_2020,Molpeceres_2024} will modify its solid-phase abundance. Furthermore, its subsequent desorption via non-thermal mechanisms will depend on the efficiency of those processes  \citep{Mennella_2004,Munoz-Caro2010,Fayolle_2011,dartois_non-thermal_2019,Kruczkiewicz_2024}.

\section{Conclusions}

We observed the cold core L694 with the IRAM 30m telescope and detected ten molecules via 11 transitions (24 when accounting for the hyperfine structure of N$_2$H$^+$), including CH$_3$OH and CO isotopologues. We then derived molecular abundance maps (300'' x 300'') for all the molecules, allowing us to study how abundances depend on the physical conditions (primarily density) and to compare with a previous study of L429-C \citep{taillard_constraints_2023}. These observations were compared to predictions from the Nautilus chemical model to constrain the time evolution of the cloud. 
Our main results are:

\begin{itemize}
    \item With the exception of CS (2-1) which exhibits two velocity components, emission spectra obtained from millimetre observations show simple velocity structure. Considering that the N$_2$H$^+$ emission is strongest at the densest part of the core and that the C$^{18}$O emission spectra widened in this area, we confirm the infalling nature of L694.

    \item All the molecules show strong depletion at the dust continuum peak, indicating a chemically advanced and dynamically active region. We find a CO depletion factor f$_{\rm CO}$ = f(X$_{\rm can}$/X$_{\rm ^{12}CO}$) of 2.23 at the dust continuum peak.

    \item We ran both static and dynamical Nautilus models to compare with our observations. Using the static models, we demonstrated that the timescale required to reproduce our gas-phase observations decreases with density, with the density itself varying by more than a factor of $\sim$ 30 across the full range explored. Overall, the abundances are reproduced within a factor of 10, which we find satisfactory.

    \item We observed a methanol gas-to-ice ratio of 0.003\% at a density of $\sim 4.3 \times 10^4$ cm$^{-3}$. We need more data points in order to recover the increasing trend with the density observed in L429-C in our previous study. This value is reproduced within a factor of 3 by the best model (with low sputtering yield).
    
    \item The physical parameters derived for the two cold cores studied (L429-C and L694) using \textit{Herschel} data show little difference. The gas phase, however, exhibits a more advanced dynamic in L694, where all the molecules show depletion profiles. This is consistent with our static models, which predicted the need for shorter times to reproduce the observations in L429-C.

    \item The dynamical SPH models exhibit a range of chemical abundances that depend on the histories of individual particles. These models fail to reproduce the abundances of most of the molecules at low density and gas-phase CO abundance at high density. The depletion trends and abundances of CS and CH$_3$OH at high densities, however, are well reproduced.  

    \item Using molecular abundances to trace the dynamical evolution of the core, we show that our dynamical models under-predict the CO gas-phase abundance at high density by at least one order of magnitude. In fact, the slow evolution of the density in these simulations results in a high depletion of CO, which is not supported by our observations. A deeper study on the parameters impacting CO depletion (sticking coefficient, density, and time) could lead to improvements. 
    
\end{itemize}

As this source is the target of a JWST GTO proposal (PID 1187), this study was conducted in preparation for the chemistry that will be revealed by the telescope. A complete ice map of the source would enable a direct comparison with the gas-phase maps we obtained and would be significantly beneficial for studying the entire ice and/or gas interface.

\begin{acknowledgements} 
We thank our referee, Olli Sipilä, for their useful comments that made the manuscript better. AT acknowledges funding from the European Research Council under the European Union's Horizon 2022 research and innovation program (grant agreement No. 101096293 SUL4LIFE). AT, VW, PG, ED, MC, and JN acknowledge the thematic Action “Physique et Chimie du Milieu Interstellaire” (PCMI) of INSU Programme National “Astro”, with contributions from CNRS Physique \& CNRS Chimie, CEA, and CNES.
\end{acknowledgements}

\bibliographystyle{aa}
\bibliography{biblio}

\clearpage

\appendix

\onecolumn

\section{Integrated intensity maps}\label{annexe-cartesintensité}

\begin{figure*}[!htp]
\includegraphics[width=0.33\linewidth]{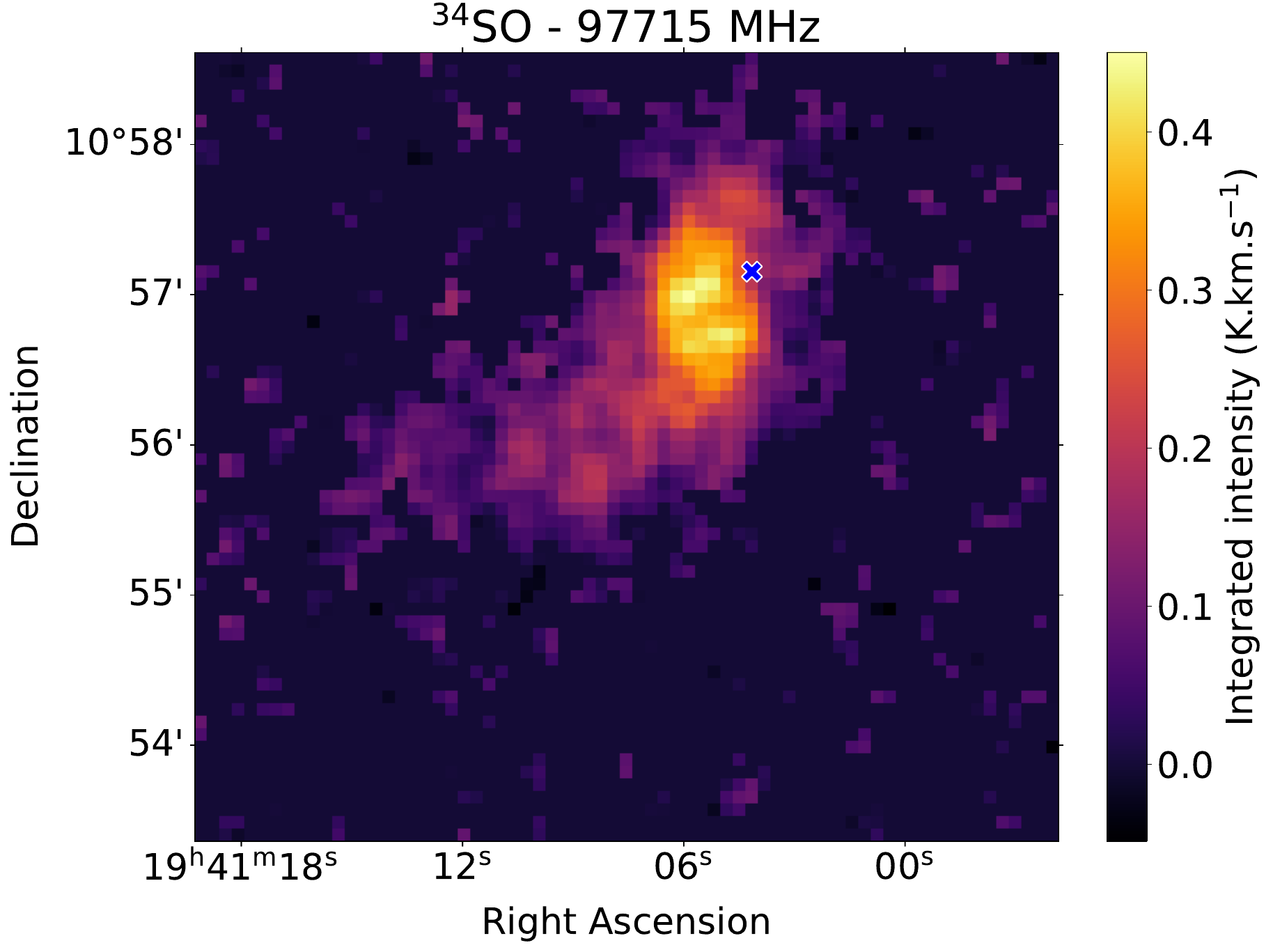}
\includegraphics[width=0.33\linewidth]{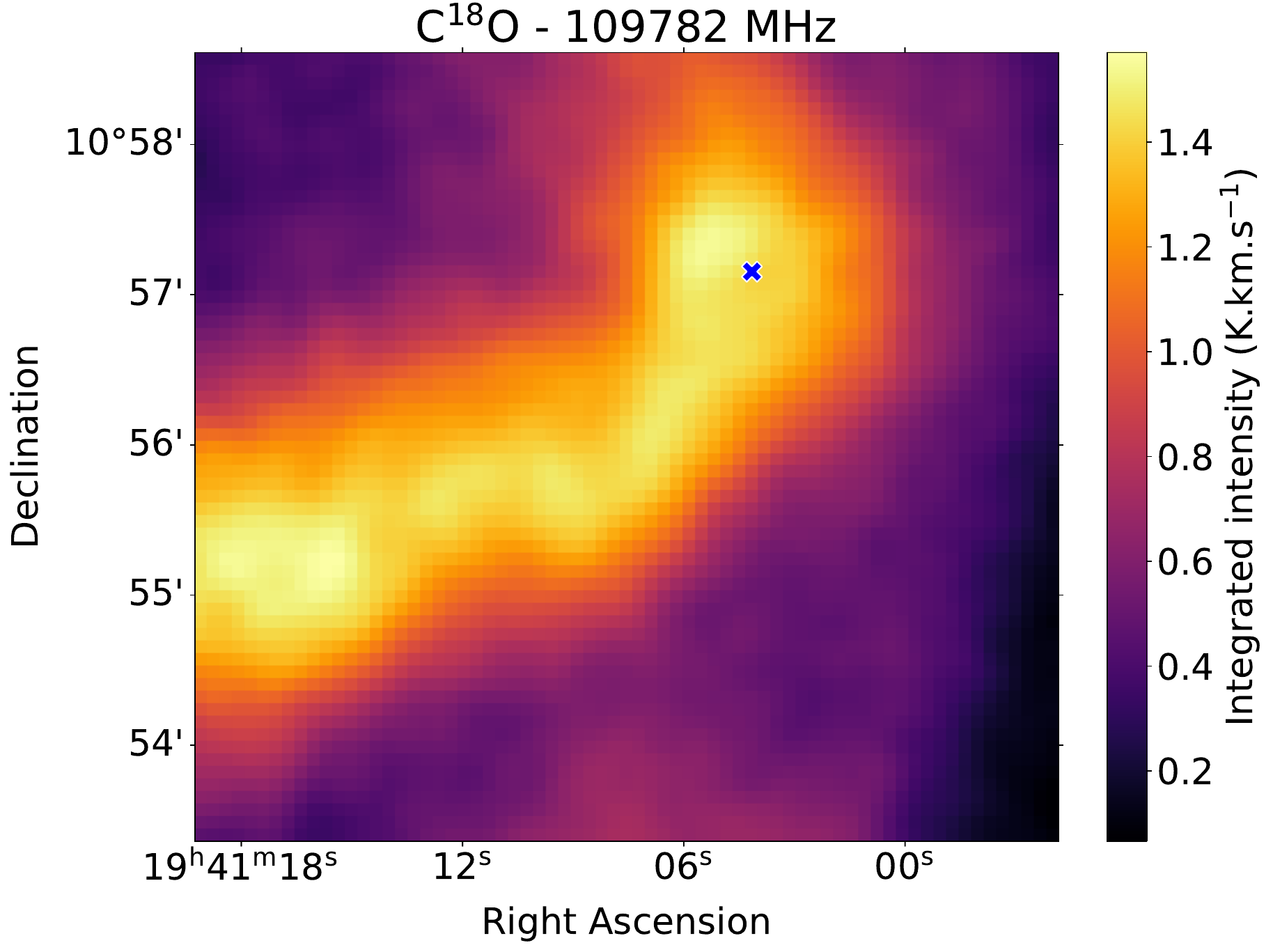}
\includegraphics[width=0.33\linewidth]{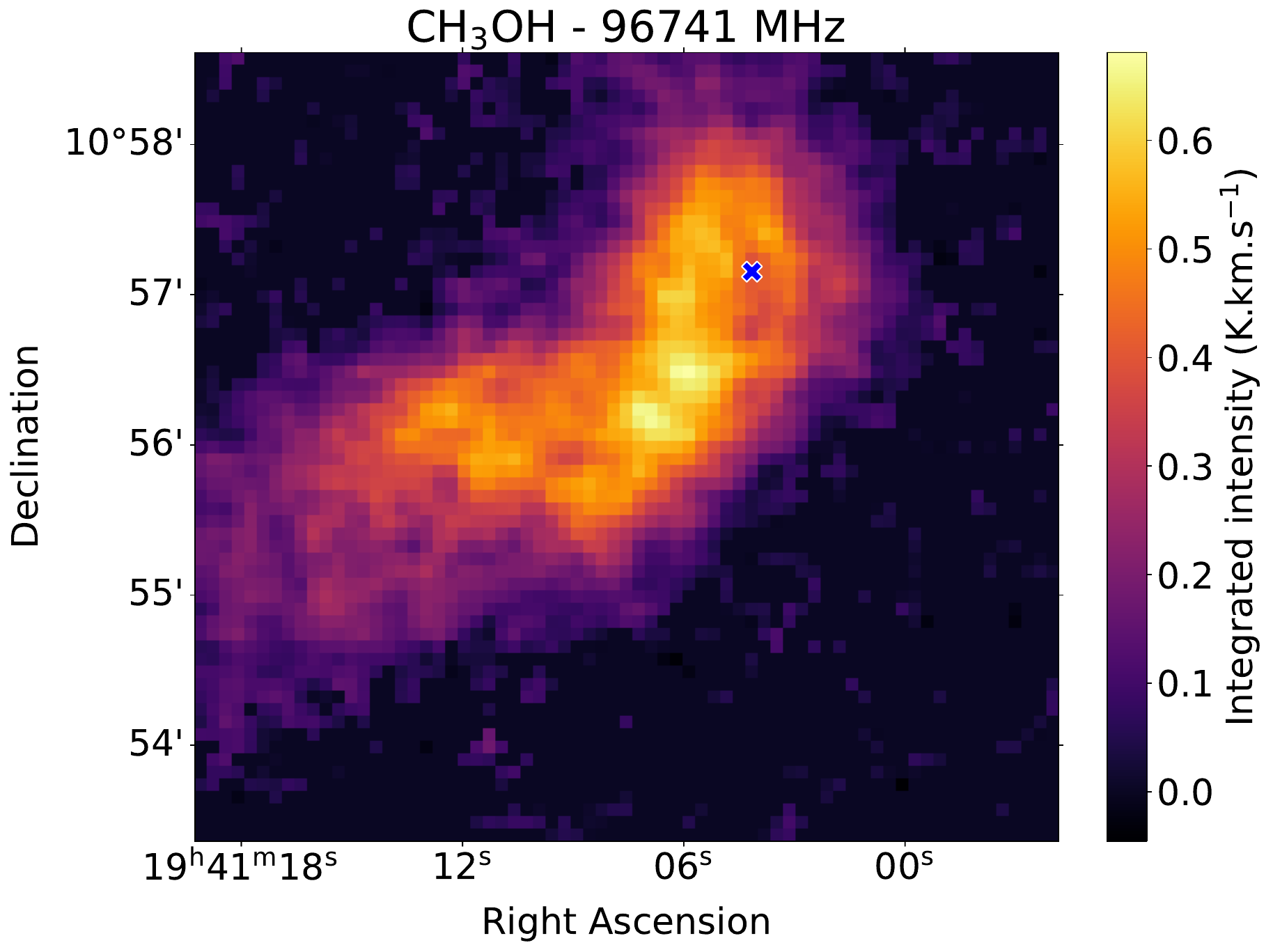}
\includegraphics[width=0.33\linewidth]{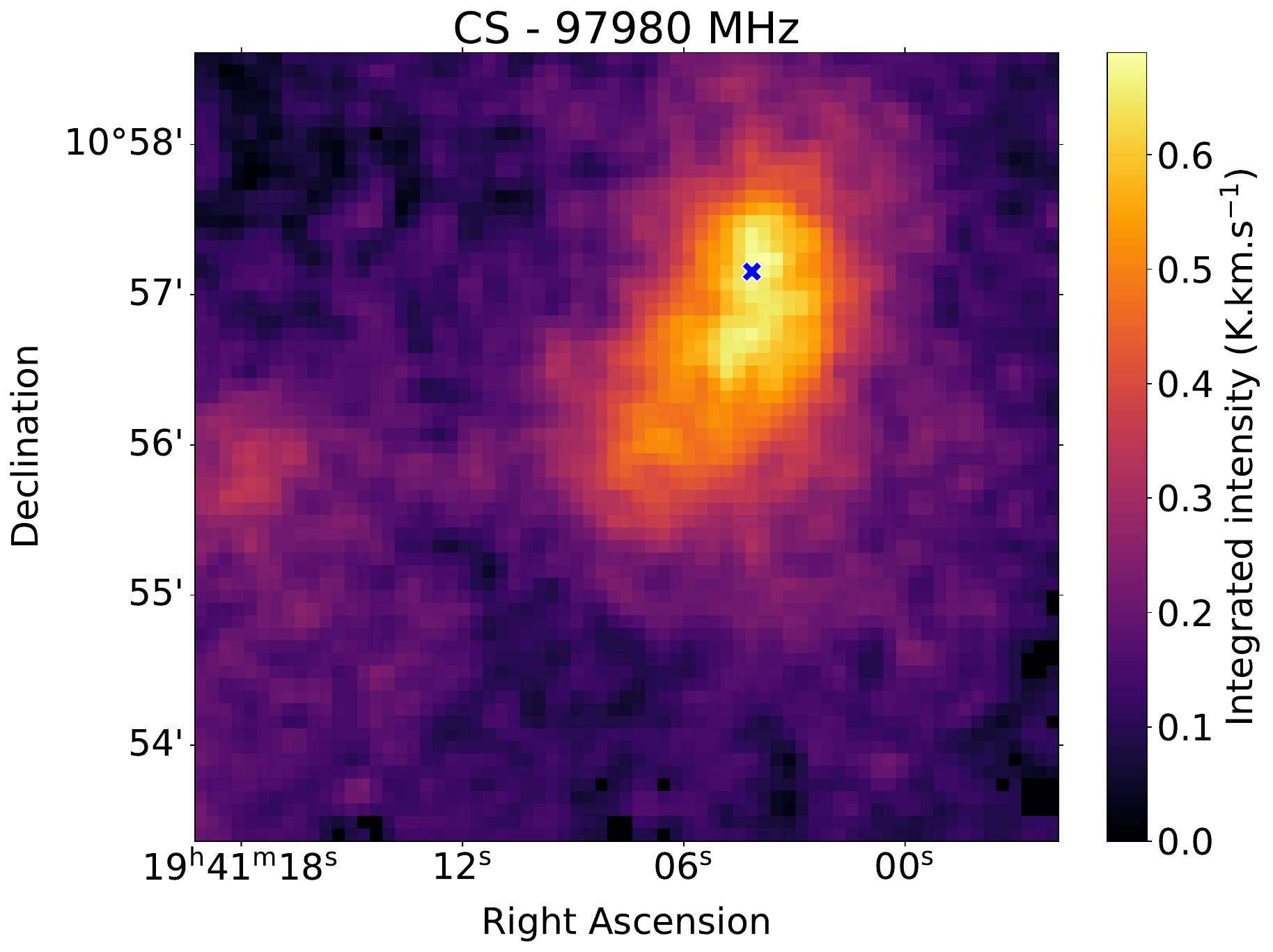}
\includegraphics[width=0.33\linewidth]{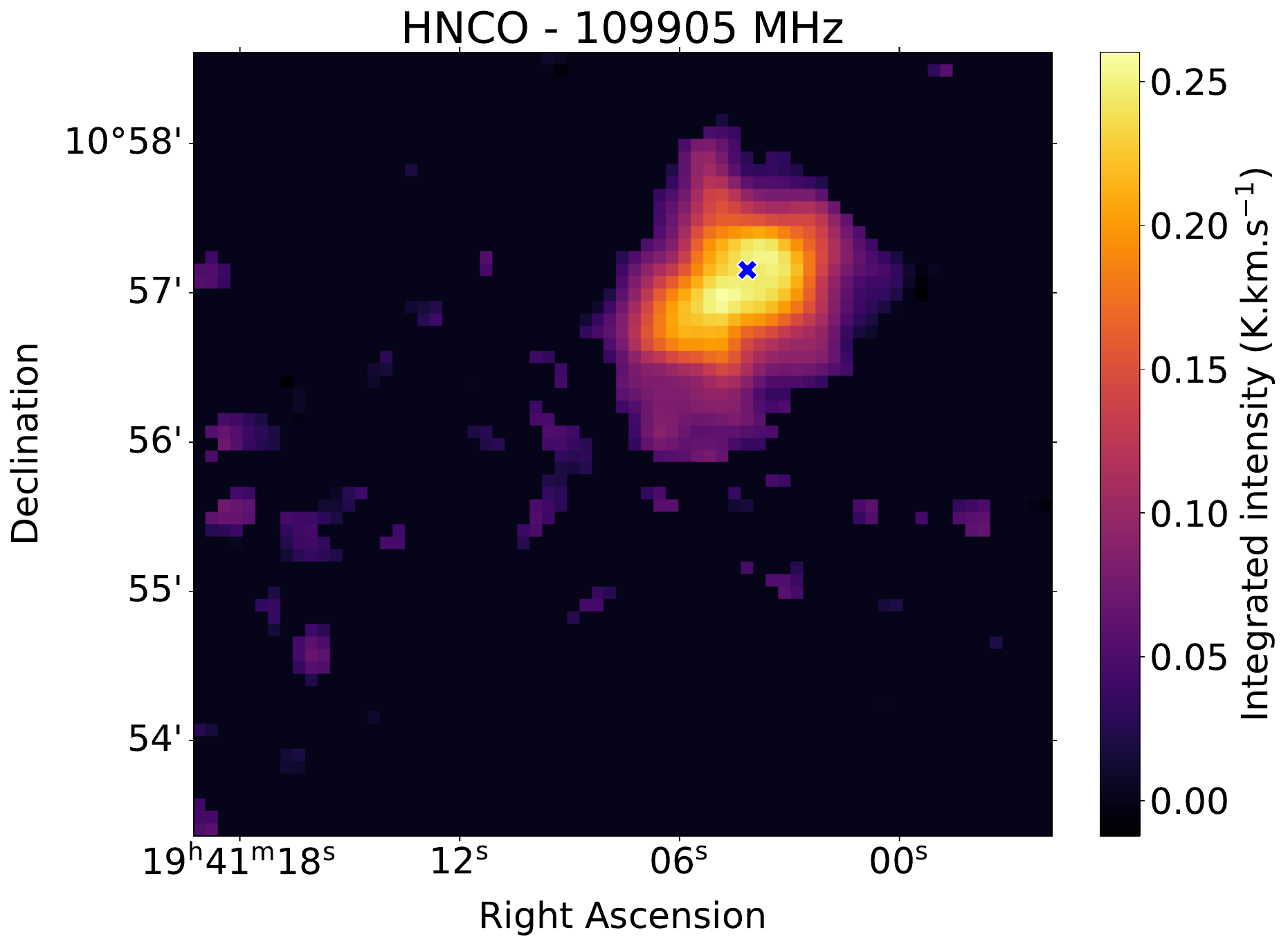}
\includegraphics[width=0.33\linewidth]{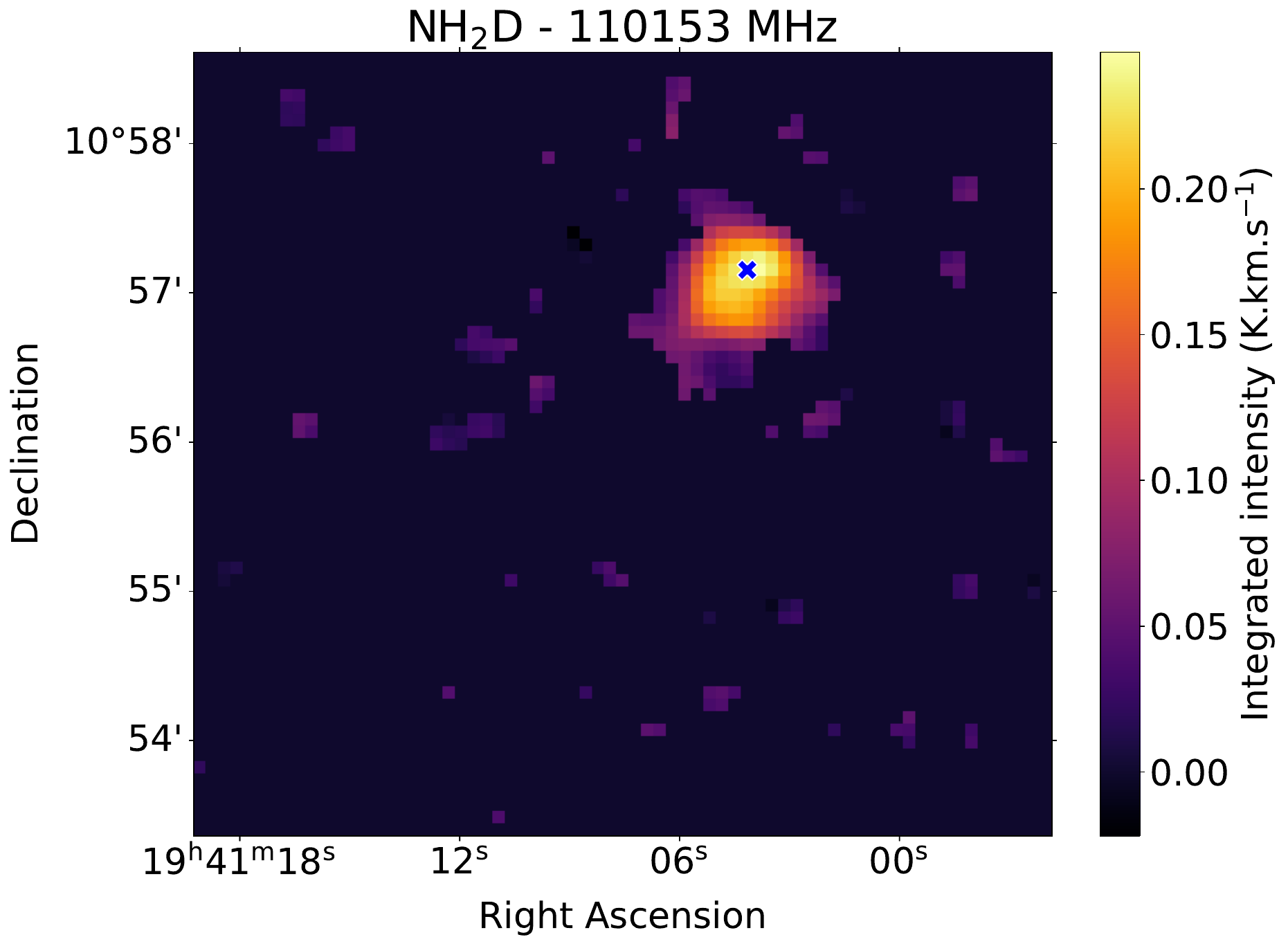}
\includegraphics[width=0.33\linewidth]{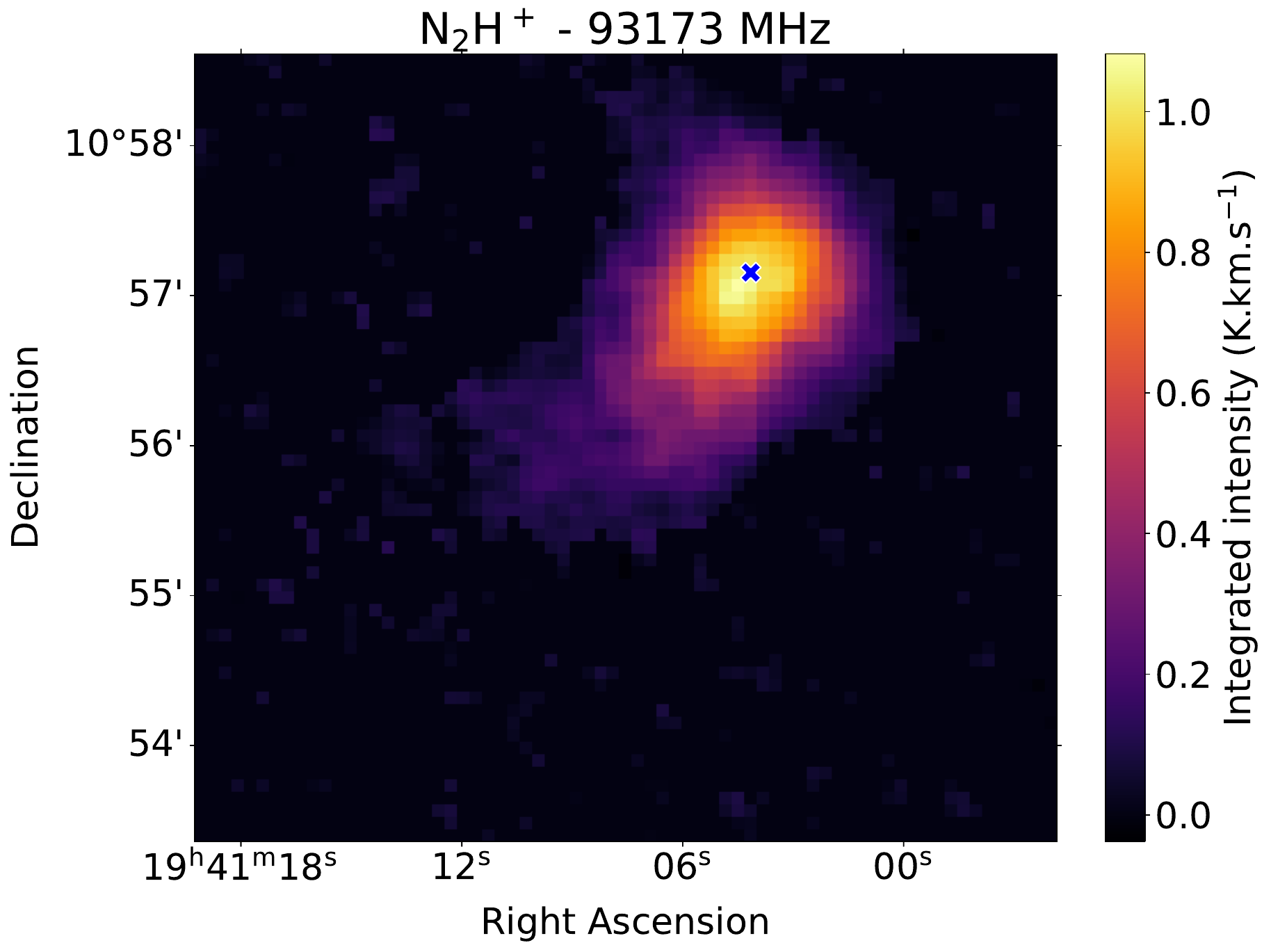}
\includegraphics[width=0.33\linewidth]{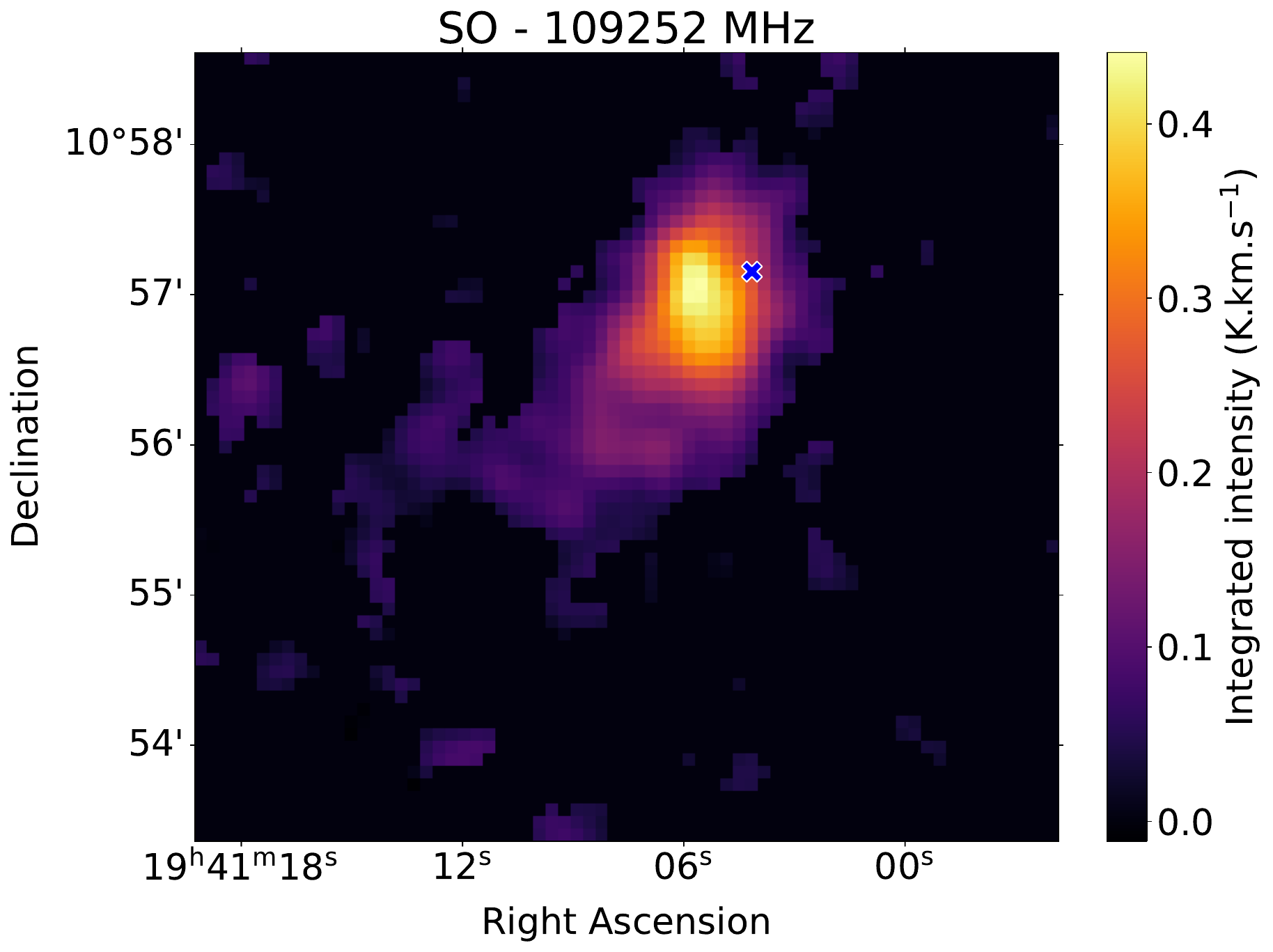}
\caption{Integrated intensity maps for each detected molecule. In the case of multiple transitions detected for one molecule, the brightest one is shown. The continuum peak is shown with a blue cross.
\label{fig:L694_intensity}}
\end{figure*}

We show the integrated intensity maps in Fig.~\ref{fig:L694_intensity} obtained for the brightest transitions of each molecule observed. These maps were obtained by integrating the velocity channels where signal is found to be $>$ 3 times the mean rms for each pixel.

\clearpage

\onecolumn

\section{Velocity channel maps}\label{velocity_channel_maps}

\begin{figure*}[h!]
     \centering
     \includegraphics[width=0.48\linewidth]{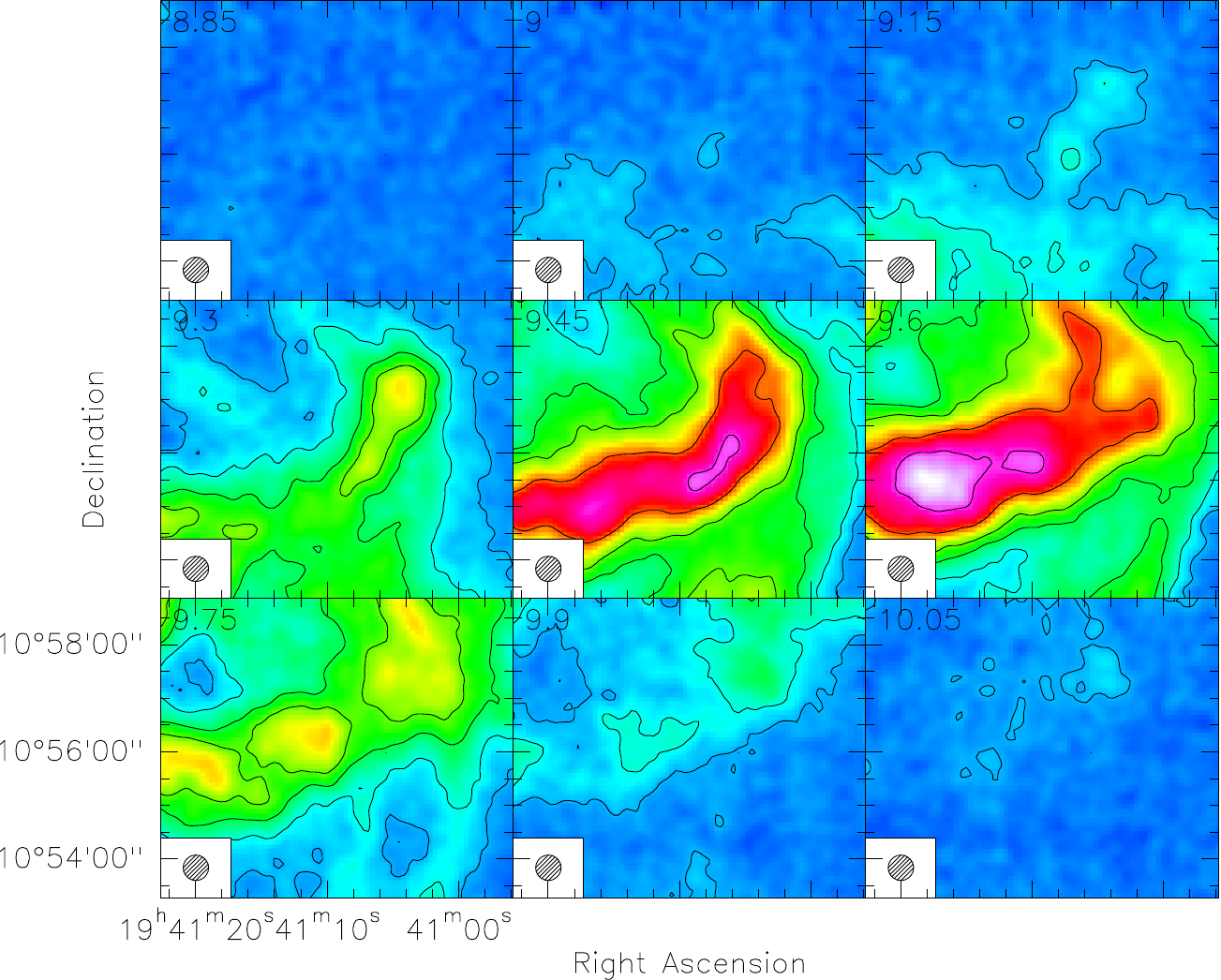}
     \includegraphics[width=0.48\linewidth]{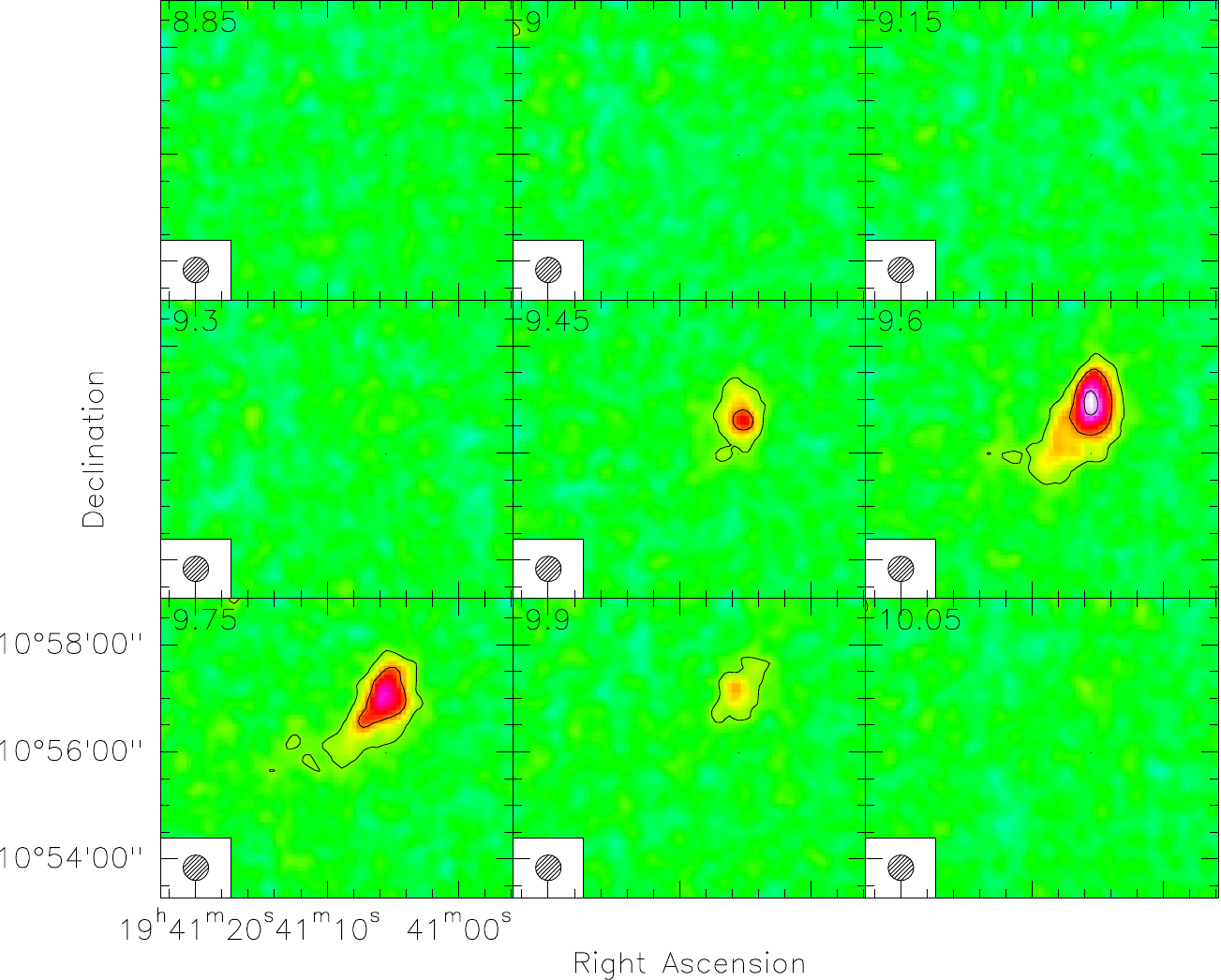}
     \includegraphics[width=0.48\linewidth]{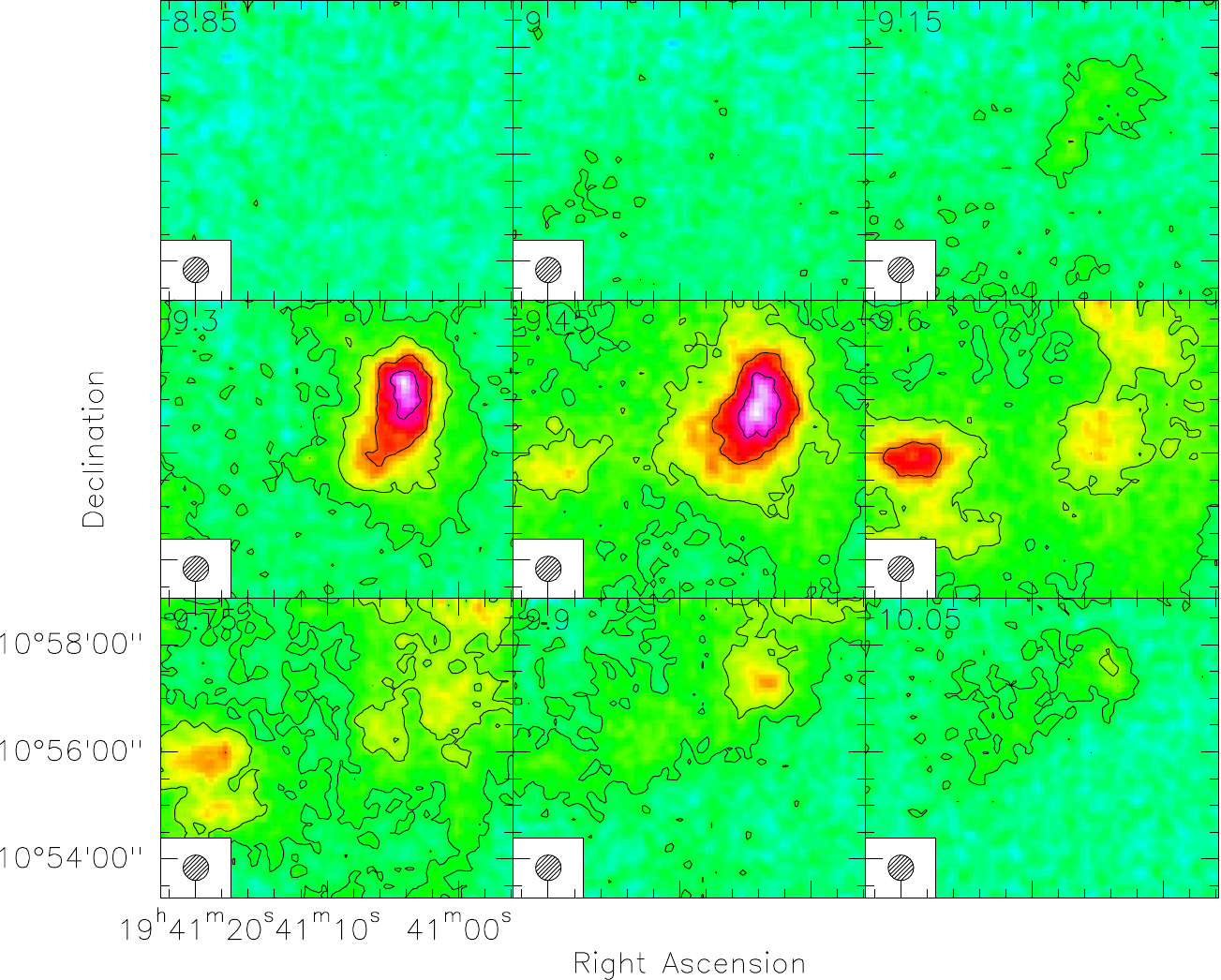}
     \includegraphics[width=0.48\linewidth]{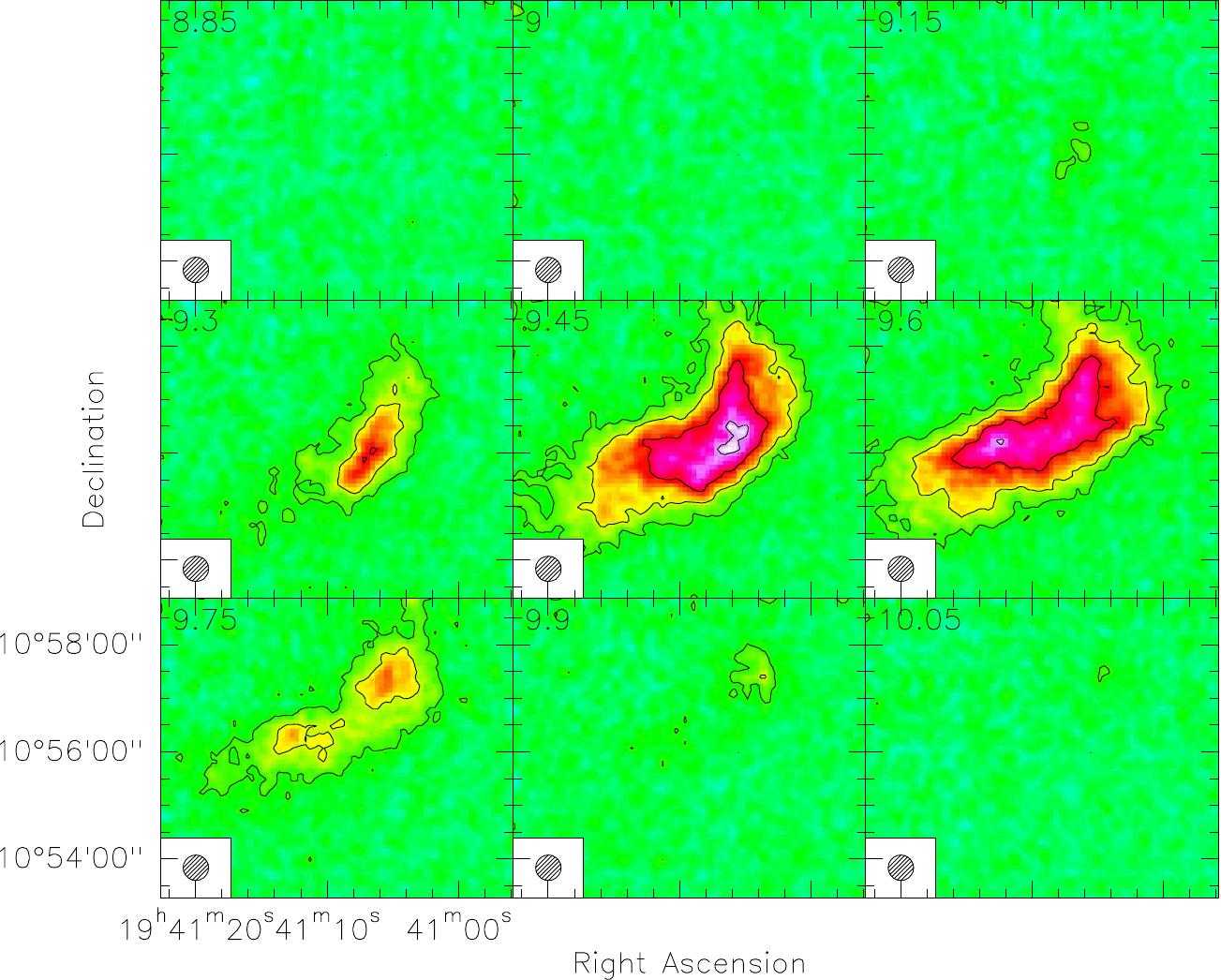}     
     
      \caption{Velocity channel maps of C$^{18}$O (109782 MHz) (top, left),  SO (99299 MHz) (top, right), CS (97980 MHz) (bottom, left) and CH$_3$OH (96741 MHz) (bottom, right), for the velocity channels from 8.85 to 10.05 km\,s$^{-1}$. The telescope beam is shown in hatched on the lower side of each panel. Contours (outer to inner) are 0.2, 0.5, 1, 1.5, 2.5, 3.5 K (T$_A^*$).
      \label{fig:l694-c18o-channel-maps} }
 \end{figure*}

We present the velocity channel maps obtained for some molecules (C$^{18}$O, SO, CS and CH$_3$OH), underlining the dynamical structure of the cloud, in Figs~\ref{fig:l694-c18o-channel-maps}.  

\twocolumn

\section{L694 physical parameters}\label{appendix_phy_param}

\begin{figure}[!h]
     \centering
     \includegraphics[width=0.5\linewidth]{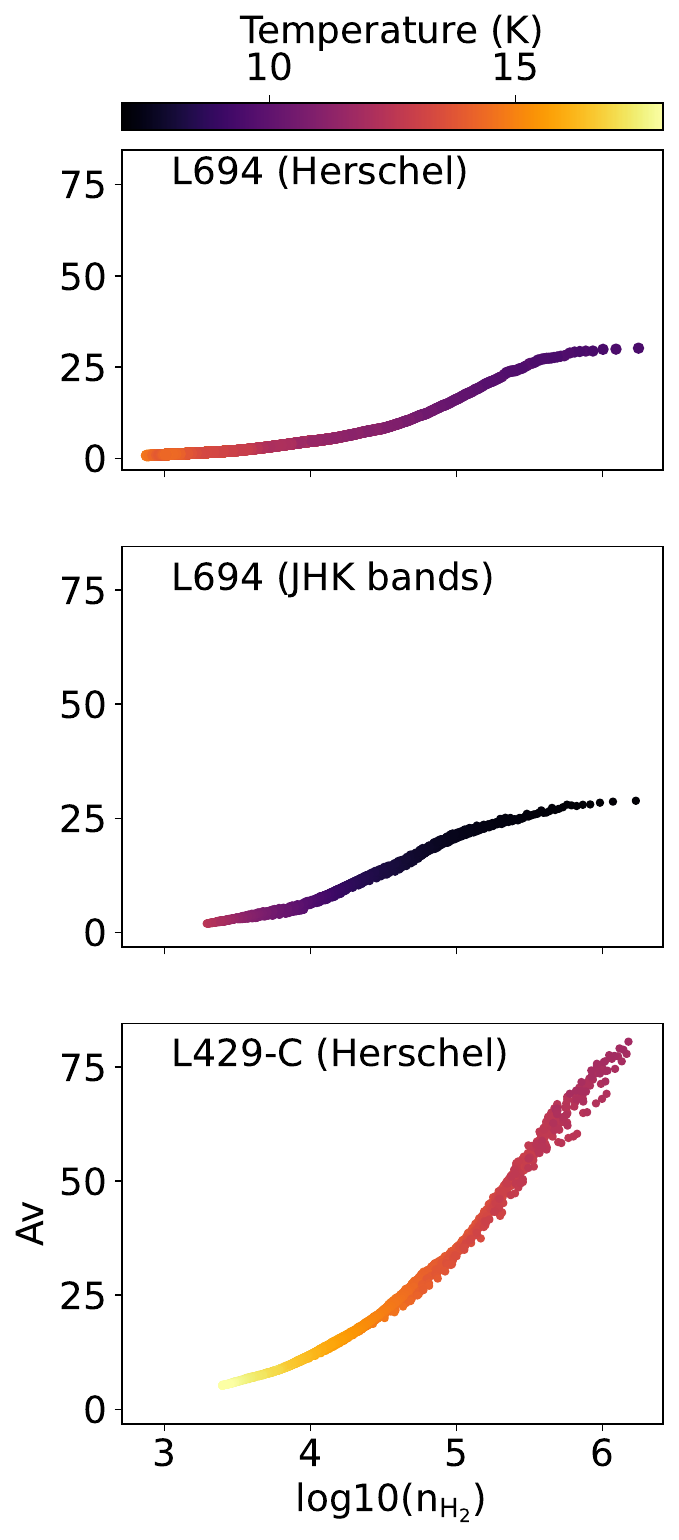}
      \caption{Av as a function of n$_{\rm H_{2}}$ (cm$^{-3}$) and in color-scale, the dust temperature (K) for, top to bottom, L694 computed from method 1, method 2 \citep{chu_constraining_2021}, L429-C \cite{taillard_constraints_2023}.   
      \label{fig:l694-parameters} }
 \end{figure}

In this section we compare how the physical parameters vary as a function of one another (Fig.~\ref{fig:l694-parameters}) for both sources (L429-C and L694). In the case of L694, both the temperatures obtained with method 1 (top) and method 2 (middle) are shown (see Sect.~\ref{sec:physical_param}). The method 2 data used to derive the physical parameters is detailed in Sect.~\ref{annexe-Chu}. 
The temperatures determined with method 2 ranges from $\sim$ 9.3 K on the core to $\sim$ 15.5 K in the outer part of the cloud for L694 (top) and from $\sim$ 11.7 to $\sim$ 18 K for L429-C (bottom). Their densities are high at the continuum peak ($> 1.5 \times 10^6$ cm$^{-3}$) and lowers on the outer part of the cloud ($\sim 10^3$ cm$^{-3}$). N$\rm_{H_2}$ was obtained from the optical depth from \textit{Herschel} with the same method as described in Appendix A in \citet{taillard_constraints_2023} and Av were computed from the conversion factor: N$\rm_H$ = 1.8 $\times$ 10$^{21}$ $\times$ Av. 
The Av varies similarly, from 25 at the continuum peak to $<$ 1 on the outer part of the cloud for L694. For L429-C, the Av is greater by a factor $>$ 2, varying from $\sim$ 9 to $\sim$ 77 at high densities, which is a result of a higher N$\rm_{H_2}$.

\section{Goodness of fit for modelled molecules}\label{goodnessoffit}

\begin{figure}
    \centering
    \includegraphics[width=0.99\linewidth]{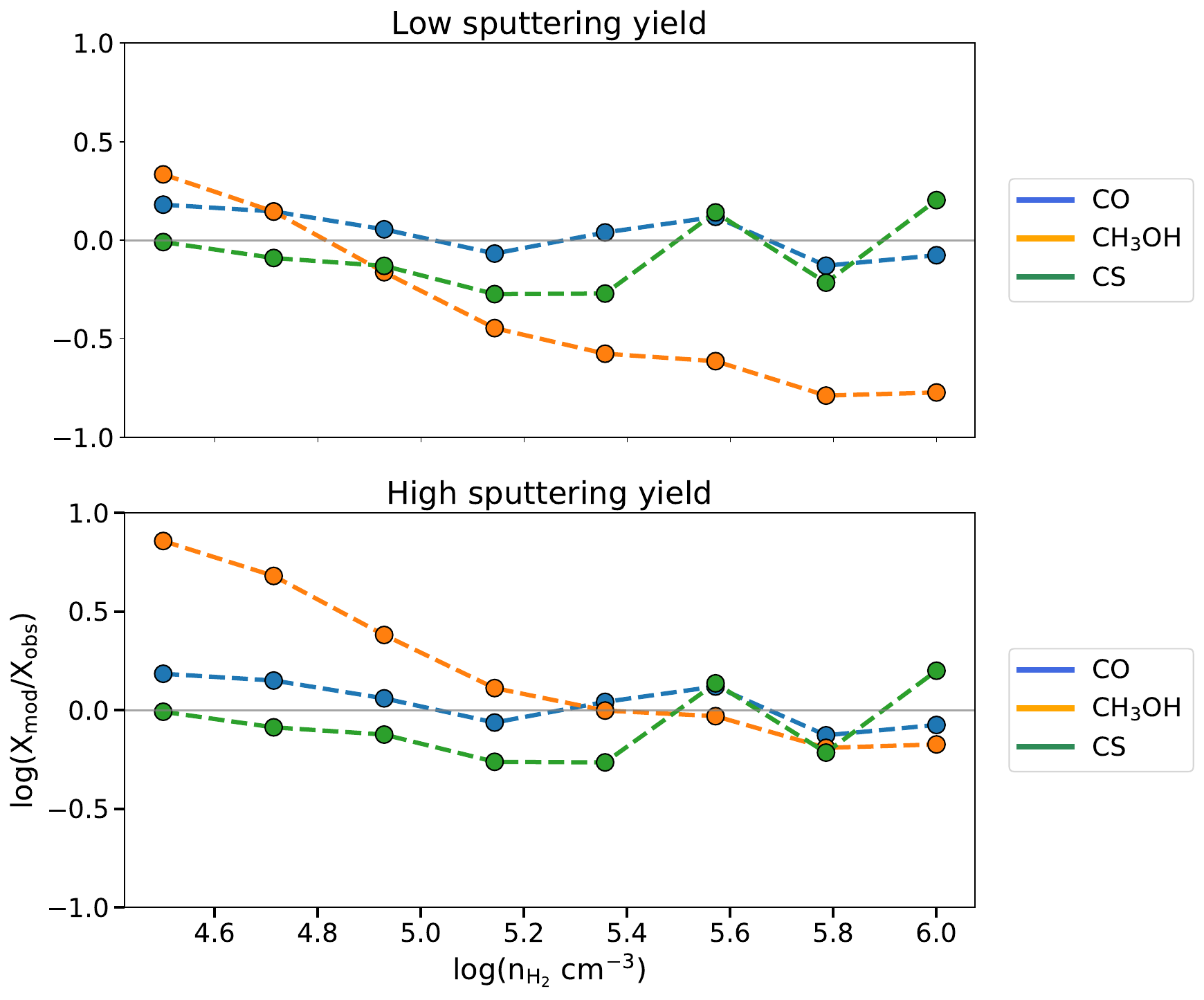}
         \caption{Ratio between modelled (X$_{\rm mod}$) and observed abundances (X$_{\rm obs}$) in the gas phase for CO, CS and CH$_3$OH as a function of density for each best time, in all four sets of models.}
    \label{fig:L694_comparison_ab_model_obs}
\end{figure}

\begin{figure}
    \centering
    \includegraphics[width=0.99\linewidth]{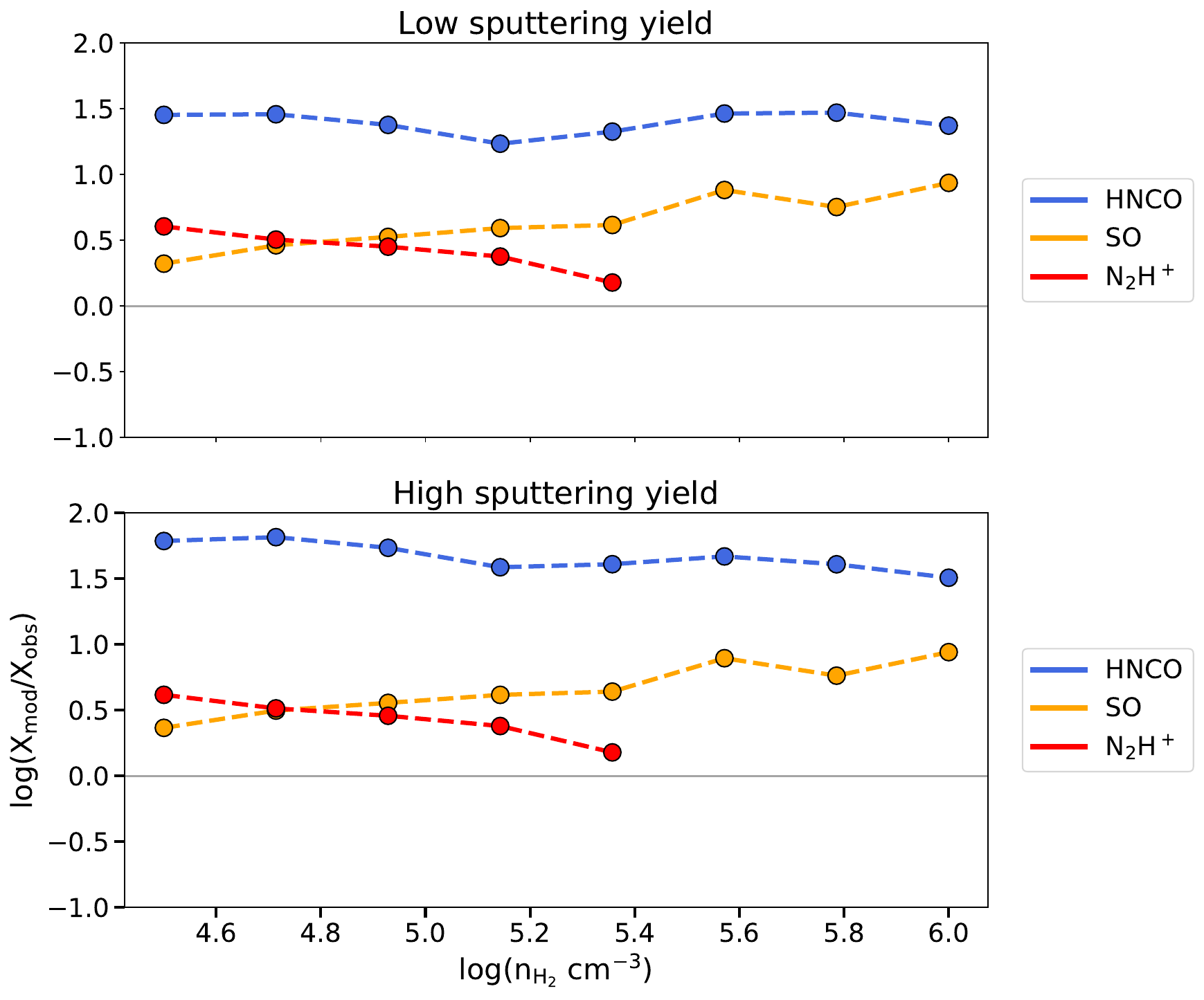}
         \caption{Ratio between modelled (X$_{\rm mod}$) and observed abundances (X$_{\rm obs}$) in the gas phase for HNCO, SO and N$_2$H$^+$ as a function of density for each best time, in all sets models.}
    \label{fig:L694_comparison_ab_model_obs_appendix}
\end{figure}

In the Figs.~\ref{fig:L694_comparison_ab_model_obs} and \ref{fig:L694_comparison_ab_model_obs_appendix}, we compare the quality of the goodness of fit by looking at the ratio between models and observations in the gas phase. Overall, all molecules are reproduced within a factor of 10 (except for HNCO, within a factor of 20) and there is no difference in chemical composition when using two different CR ionisation yields. 
CO and CS very are well reproduced in all models.
For CH$_3$OH, the models with a low sputtering are underproducing (by a factor 5) the molecule as density increases while in the high sputtering ones, the low density abundances are overproduced (by a factor $>$ 6). It is however the models with low sputtering that are the closest to the observed methanol gas-to-ice ratio (See Sect.~\ref{staticmodels-results}).
SO is over-produced in all models, from a factor 3 to almost 10 at high density. 
N$_2$H$^+$ is over-reproduced by a factor $\sim$ 2 on the entire range of densities where it is detected. 
HNCO is the molecule with the worst $\rm X_{mod}/X_{obs}$ ratio, being over-produced by a factor $\sim$ 30 on all the density range.
In \citet{wirstrom_search_2016}, the authors measured a O$_2$ abundance in L694 of $<$ 1.6 $\times 10^{-7}$ on the maximum of the dust continuum. With Nautilus, we predict an abundance of $\sim$ 5 $\times 10^{-7}$, which is over-produced by a factor of 3.

\section{Deriving abundance maps using background stars observations (``method two'')}\label{annexe-Chu}

\begin{figure}
    \centering
    \includegraphics[width=0.75\linewidth]{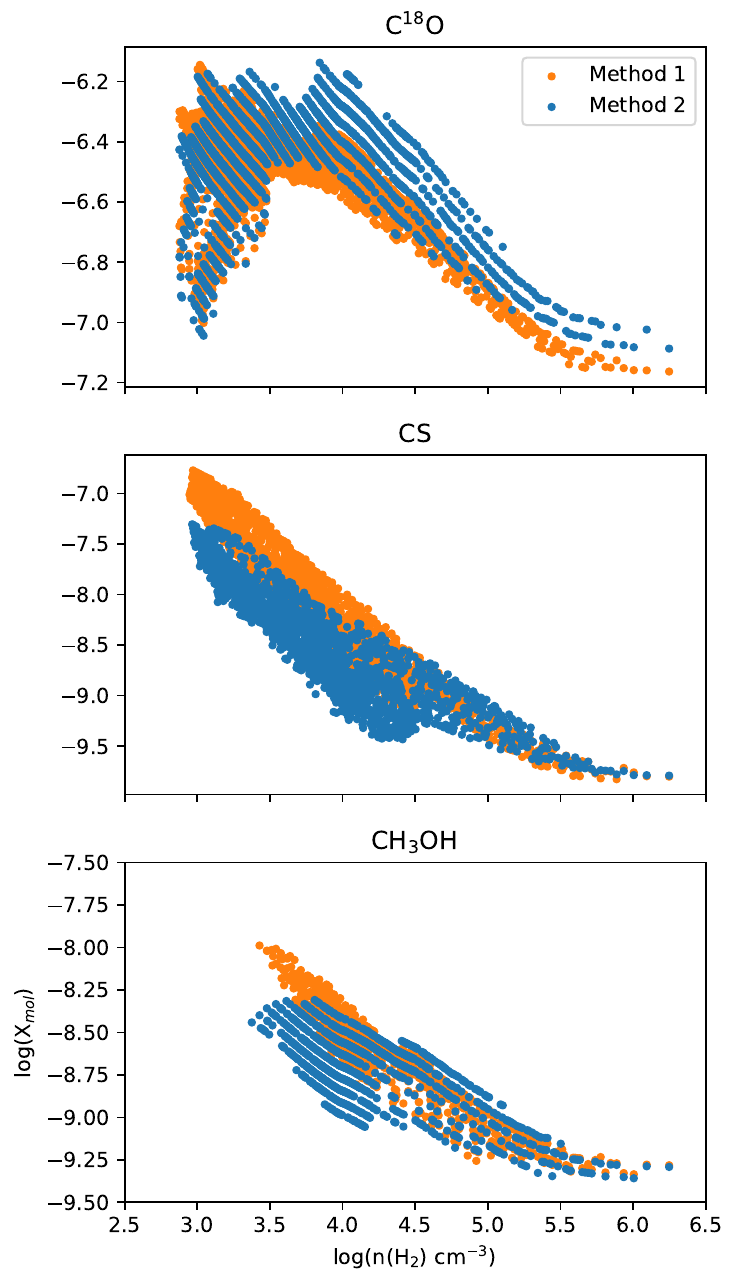}
    \caption{Comparison between the gas-phase abundances for C$^{18}$O, CS and CH$_3$OH between method 1 (in orange) and method 2 (in blue). }
    \label{fig:comp_laurie_nous}
\end{figure}

We used the background star data provided by \citet{chu_constraining_2021} to construct physical parameters maps with a different method other than method 1. In this study, the authors have determined the extinction from background stars photometry \citep[see section 4.1 of][]{chu_constraining_2021}, and obtained a value of Av for each position of stars they observed. Using a similar smoothing method as they did, we managed to derive an almost identical extinction map as theirs (Fig. 1 in their study). 
The Av determined on the dust continuum peak has bigger uncertainties, as it is too dense to observe any background stars and we conveyed values by smoothing everything with a Gaussian kernel. The authors of the study also noted that in the region of highest extinction, the resolution under-samples the data, meaning that the Av determined in this area (shown as black pixels in their Figure 1) is interpolated from the three closest stars at most. 
Once we obtained the Av map, we derived the H$_2$ column density map by using the following equation \citep{predehl_x-raying_1995,ryter_interstellar_1996,olofsson_extinction_2010}: $\rm N_{H_2} = 1.8 \times 10^{21} / 2$. 
The H$_2$ column density ranges from $\sim$ 2 $\times 10^{21}$ to 2.6 $\times 10^{22}$ cm$^{-2}$. 
From N$\rm_{H_2}$, we derive the H$_2$ volume density n$\rm_{H_2}$ by using the same method as our previous work \citep[derived from][]{bron_clustering_2018}. This gives us a volume density ranging from $\sim 1.9 \times 10^3$ to $\sim 1.7 \times 10^7$ cm$^{-3}$. 
Lastly, to obtain the dust temperature, we used the same formula as described in section~\ref{nautilus-section} from the parametrisation derived in \citet{hocuk_parameterizing_2017}.
The temperature obtained range from $\sim$ 7 K to $\sim$ 12.5 K, which is the coldest of the temperatures presented in Figure~\ref{fig:l694-parameters}, plotted as the colour-map of the density as a function of Av. 
With these physical parameters, we computed the molecular column densities and then abundances for all the detected molecules. We found that the derived molecular abundances are similar (within a factor of 3) to the ones we obtained with method 1. We show in Figure.~\ref{fig:comp_laurie_nous} the different abundances of CO, CS and CH$_3$OH obtained with the two methods.

\begin{figure}
    \centering
    \includegraphics[width=0.85\linewidth]{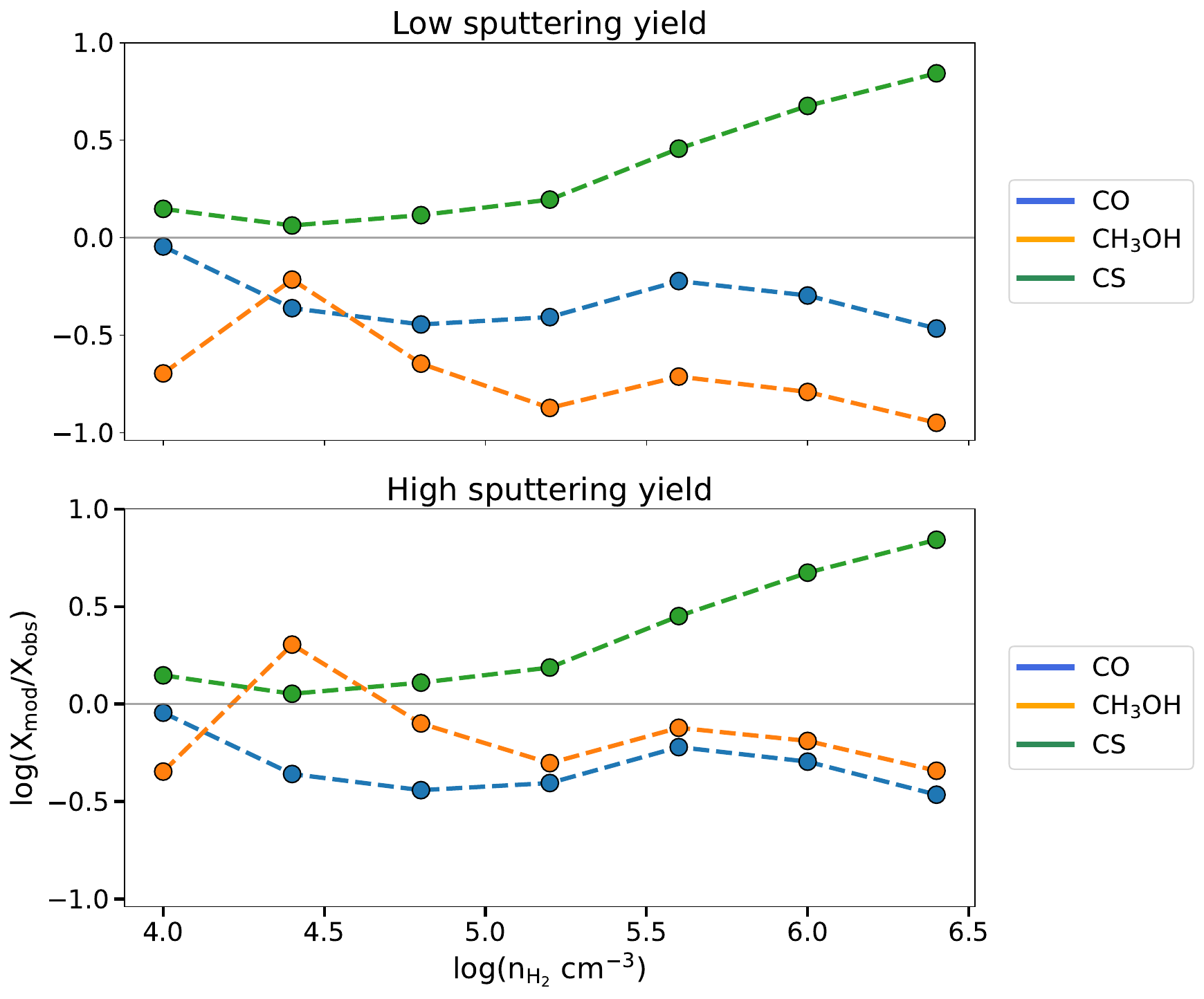}
         \caption{Ratio between modelled (X$_{\rm mod}$) and observed abundances (X$_{\rm obs}$) in the gas phase for CO, CS and CH$_3$OH as a function of density for each best time, in all four sets of models for method 2.}
    \label{fig:Laurie_model_vs_obs}
\end{figure}

We also ran the static chemical models described in section~\ref{static-models} using these new set of physical parameters and compared with the observed abundances derived using these parameters. 
The best time also varies linearly and are slightly smaller than the one we derived using method 1 ranging from  1 $\times 10^6$ to 7.7 $\times 10^3$ yr for all set of parameters. 
The goodness of fit for gas-phase species is slightly worse than with method 1 but all the predictions are still fitting within a factor 10. In Figure.~\ref{fig:Laurie_model_vs_obs} we show the goodness of fit for CO, CS and CH$_3$OH.
CO is reproduced within a factor 3 no matter the set of models or density. CS is reproduced within a factor of two at density $<$ 10$^5$ cm$^{-3}$ and the agreement worsens as density increases.
SO at low density is 10 times under-produced compared to observations but the ratio X$\rm_{mod}$/X$\rm_{obs}$ increasing with density and is overproduced at higher density within a factor 3.
Methanol is under-produced (up to a factor 10) in low sputtering yield models at densities higher than 10$^{5}$ cm$^{-3}$. In high sputtering yield model, the fit is bad at densities lower than 10$^{5}$ cm$^{-3}$ and then is well reproduced at higher densities (within a factor 3). 
N$_2$H$^+$ is reproduced the same than what is shown in Fig.~\ref{fig:L694_comparison_ab_model_obs_appendix}.
HNCO is better reproduced within a factor 10, favoured by the cold chemistry in these physical parameters as compared to the models presented in this study. 
CO ices and methanol ices are predicted the same and the CH$_3$OH gas-to-ice ratio is identical, both in behaviour and values in all sets of models.
Overall, a few species are strongly impacted by low temperature chemistry (CS, SO) while others (CH$_3$OH, CO, HNCO) have their behaviour changed compared to method 1 but are still well reproduced. 

\end{document}